\documentclass[twocolumn,floats,pra,eqsecnum,showpacs]{revtex4-1}

\usepackage{amsmath}
\usepackage{amsfonts}

\usepackage{tikz}

\newcommand {\be}{\begin{equation}}
\newcommand {\ee} {\end{equation}}
\newcommand {\bea}{\begin{eqnarray}}
\newcommand {\eea} {\end{eqnarray}}
\newcommand{\non}{\nonumber}

\begin{document}


\title{Complexity as information in spin-glass Gibbs states and metastates:\\
upper bounds at nonzero temperature and long-range models}
\author{N. Read}
\affiliation{Department of Physics, Yale University, P.O. Box 208120, New Haven, CT 06520-8120\\
Department of Applied Physics, Yale University, P.O. Box 208284, New Haven, CT 06520-8284}
\date{June 6, 2022}

\begin{abstract}
In classical finite-range spin systems, especially those with disorder such as spin glasses, a low-temperature Gibbs state may be 
a mixture of a number of pure or ordered states; the complexity of the Gibbs state has been defined in the past 
roughly as the logarithm of this number, assuming the question is meaningful in a finite system. As non-trivial mixtures of pure states
do not occur in finite size, in a recent paper [Phys. Rev. E {\bf 101}, 042114 (2020)] H\"oller and the author introduced 
a definition of the complexity of an infinite-size Gibbs state as the mutual information between the pure state and the spin configuration 
in a finite region, and applied this also within a metastate construction. (A metastate is a probability distribution on Gibbs states.)
They found an upper bound on the complexity for models 
of Ising spins in which each spin interacts with only a {\em finite} number of others, in terms of the surface area of the region, 
for all $T\geq 0$. In the present paper, the complexity of a metastate 
is defined likewise in terms of the mutual information between the Gibbs state and the spin configuration. Upper bounds are found 
for each of these complexities for general finite-range (i.e.\ short- or long-range, in a sense we define) mixed $p$-spin 
interactions of discrete or continuous 
spins (such as $m$-vector models), but only for $T>0$. For short-range models, 
the bound reduces to the surface area. For long-range interactions, the definition 
of a Gibbs state has to be modified, and for these models we also prove that the states obtained within the 
metastate constructions are Gibbs states under the modified definition. All results are valid for a large class of disorder distributions. 
\end{abstract}


\maketitle

\section{Introduction}
\label{intro}

One of several uses of the term ``complexity'' in the physical and mathematical sciences refers to
the decomposition of an equilibrium (Gibbs) state in a short-range classical statistical mechanics system as a mixture of 
ordered or ``pure'' Gibbs states \cite{georgii_book,simon_book,bovier_book}, or to the similar decomposition 
of stationary (in time) statistical states of a corresponding dynamical system (i.e.\ with the same Hamiltonian) 
into ergodic components \cite{liggett_book}. (Indeed, the Gibbs states are stationary, and the pure states are ergodic.) 
Each weight in the mixture can be viewed as the probability for the system to be in the corresponding pure state, and 
Palmer \cite{palmer} suggested the name complexity for the entropy of the set of weights, and thus for (roughly speaking) 
the logarithm of the number of pure or ergodic components in such systems. He also suggested that, in some systems such 
as short-range spin glasses (SGs), the complexity could be extensive, that is, proportional to system size.

There are two issues with the last proposal. One is that it had to refer to a system of finite size 
(i.e.\ having a finite number of degrees of freedom), and typical finite-size models have only a unique pure or ergodic 
state; a nontrivial pure- or ergodic-state decomposition can arise only in an infinite-size version 
\cite{georgii_book,simon_book,bovier_book,liggett_book}. Glossing over this fact for a moment, and treating the pure 
states as somehow defined approximately for finite size, 
van Enter and van Hemmen \cite{vEvH} showed that the complexity of pure states in short-range models cannot be 
extensive. (In contrast, ordinary entropy differs on both counts: it is well-defined in the finite-size case, and extensive.)

In their deep studies of short-range finite-dimensional classical SGs which used rigorous notions of Gibbs and 
pure states in strictly infinite size \cite{ns96b,ns_rev}, Newman and Stein (NS) frequently emphasized the idea of 
restricting the Gibbs probability
distribution for the spins to only the spins in a finite-size region or ``window'', such as a hypercube $\Lambda_W$ of side $W$. 
The entropy of such a distribution is finite even when the total system size is infinite. Further, if for example one considers
different finite sizes of the system, with the interactions among the spins inside and near the window fixed, then at 
zero temperature different ground states may be seen in the window. It is clear, for example, in a model with 
interactions between only nearest-neighbor Ising spins that, except in certain degenerate models, the logarithm of the number 
of such ground states cannot be greater than a constant times the surface area $\sim W^{d-1}$ of the window. This is 
more restrictive than simply being subextensive, even if extensive is taken to mean $\sim W^d$. Because, in such systems, 
ground states are the zero-temperature version of pure states, this raises the expectation that the same bound should also 
hold for some notion of the complexity of a Gibbs state as seen in a window at non-zero temperature. We are not aware of
a precise formulation and proof of such a result in print until recently. 

In previous work, H\"oller and the author \cite{hr} (we refer to this paper as HR), building on the notions of complexity 
from Refs.\ \cite{palmer,vEvH} but using the spin distribution restricted to a finite window $\Lambda_W$, 
arrived at a definition of complexity, relative to the window, of an infinite-size Gibbs state (i.e.\ of its pure-state decomposition), 
as the {\em mutual information} between the spins in the window and the pure states. Here, the mutual information represents
the average amount of information about which pure state the system is in that is obtained by an observation of the spins in the window. 
This definition has several desirable properties, and in some situations it reduces to the entropy of the set of weights of pure states
as $W\to\infty$. For Ising spins and nearest-neighbor interactions it was easy to show \cite{hr} that it is less 
than a constant times the surface area $\sim W^{d-1}$ of the window, for any temperature $T\geq 0$. 

While the definition of complexity as mutual information in HR is very general, the upper bound on it obtained there (essentially 
by counting distinct spin configurations on the boundary of the window) is limited 
in scope in two ways, as just mentioned: (i) the bound does not cover the case of models in which each spin interacts directly 
with infinitely many others, but which, because the interactions fall off sufficiently rapidly with distance, nonetheless behaves
like a short-range model, with well-defined Gibbs states (we say models in this larger class are of finite range); (ii) while the bound 
extends trivially to other models of spins that take a finite number of distinct values, it diverges for continuous spins, such 
as vector spins with rotation-invariant interactions.  

In Section \ref{sec:compl} of this paper, we prove upper bounds on the disorder-average of the complexity that overcome 
both of these limitations for a broad class of models, but only for the case of strictly positive temperature. For interactions 
involving mean-zero bonds $J_{ij}$ between the pair of spins at $i$, $j$, we prove that the disorder 
average of the complexity (mutual information) of a Gibbs state in a SG is bounded at any $T>0$ by
\be
\frac{1}{T^2}\sum_{\{i,j\}:i\in \Lambda_W,j\not\in\Lambda_W} {\rm Var}\, J_{ij},
\ee
where ${\rm Var}\,J_{ij}$ is the variance of $J_{ij}$. The bound extends naturally to models 
with interactions among sets of $p$ spins, instead of $p=2$. [For these, the precise statements are in Sec.\ \ref{subsubcomp},
inequality (\ref{boundfinal}), and in most general form (\ref{boundfinal'}), which includes non-SGs and non-disordered models.] 
It applies to continuous spins as well as 
to discrete, provided the spins are vectors of magnitude $\leq 1$, and it reduces to the surface area $\sim W^{d-1}/T^2$
for nearest-neighbor interactions. For the one-dimensional power-law model \cite{kas}, in which $i$, $j\in{\bf Z}$ 
and ${\rm Var}\,J_{ij}=|i-j|^{-2\sigma}$ ($\sigma>1/2$), it gives a bound $\propto W^{2-2\sigma}/T^2$ for 
$1/2<\sigma<1$, and constant$/T^2$ for $\sigma>1$.  
While this bound may not be useful if one wishes to take $T\to0$ at fixed $W$, nonetheless for any fixed $T>0$ it gives a bound on 
the asymptotic growth of the average complexity as $W\to\infty$. 

The infinite-size Gibbs states that we consider in these results are obtained from taking a thermodynamic limit. As the limit may not
exist directly, it is necessary to utilize a metastate, a probability distribution on infinite-size Gibbs states for given bonds 
\cite{ns96b,ns_rev,aw}; a metastate carries information about finite-size systems. The same bound applies to the 
disorder-average both of the complexity of a Gibbs state sampled from the metastate, and also of the complexity of the 
metastate-average state (MAS), which is itself a Gibbs state. We further define a concept of the {\em complexity of the metastate} itself, 
again relative to the window. This does not refer to a pure-state decomposition, but describes the logarithm of the 
number of Gibbs states over which the metastate ranges; it vanishes for a trivial metastate (i.e.\ one supported on a single Gibbs state
for the given bonds). 
Its disorder average is subject to 
the same bound too. The three complexities are related by a simple formula: the complexity of the MAS is the sum of the
other two.  

In long-range models, such as cases $\sigma<1$ in the one-dimensional power-law models, from a rigorous point of view, 
questions arise even about the definition of Gibbs states \cite{gns}. As those models are of interest in the current work, we 
address those problems as well. It turns out that the concept of relative 
entropy, to which mutual information is closely related (and which also appears in the proofs of the bounds already discussed),
is very useful in proving the existence of metastates and the nature of the Gibbs states in the long-range models, and the methods 
used are needed for proving the bounds on complexity in these models as well. For these reasons we also include these results here, 
in the form of Appendix \ref{app:gibbs}. We also use similar methods in a proof in Appendix \ref{app:trivi} that there 
is a unique Gibbs state at $T>0$ in the short-range case $\sigma>1$ of the one-dimensional power-law models. 

In Sec.\ \ref{subsubdisc} we give a final discussion of the relation of the results to the problems of SGs.

\section{Spin-glass models}
\label{sgmodels}

We begin by detailing the SG models we have in mind; this section can be skipped or skimmed by knowledgeable readers.
(For a general reference, see e.g.\ Ref.\ \cite{cg_book}.) The notation $\Lambda$, $\Lambda'$, \ldots, will stand for sets
with elements $i$. The basic general form of a SG Hamiltonian, due to 
Edwards and Anderson (EA) \cite{ea}, is
\be
H=-\!\sum_{\{i,j\}\in{\cal E}}J_{ij}s_is_j -h\sum_i s_i,
\ee
where at present $s_i=\pm 1$ for all sites (vertices) $i$ in some (possibly infinite) index set $\Lambda$ are Ising spins, 
and $\cal E$ is a set of edges, that is unordered pairs $\{i,j\}$ of sites $i$, $j$ in $\Lambda$; the real 
numbers $J_{ij}$ are random variables called bonds, and $h$ is a magnetic field. (To be clear, the symbol $\Lambda$ 
for a set of $i$ will not always denote the system as a whole, but here it does.) At present, this expression is formal,
that is we do not yet concern ourselves with convergence of the sum. Unless stated otherwise, we 
assume the bonds (and their generalizations $J_X$ below) are independent random variables with mean zero and finite variance, 
and at first we can assume they are Gaussian. (A random variable with mean zero will be termed ``centered''.) We write $s$ for 
$(s_i)_{i\in\Lambda}$ 
and $J= (J_{ij})_{\{i,j\}\in{\cal E}}$. We also write $s|_{\Lambda'}$ for the restriction $(s_i)_{i\in\Lambda'}$ of $s$ to 
its values on a subset $\Lambda'\subseteq \Lambda$, and similarly for $J$. (Similar notation will be used for other indexed sets
and their restrictions.)
We can view a configuration $s$ as a function 
from $\Lambda$ into $\{-1,+1\}$, so  $s\in S_\Lambda =\{-1,+1\}^\Lambda$, the set of all such functions; for 
$\Lambda={\bf Z}^d$, we write $S$ for $S_\Lambda$. The probability distribution on $J$ 
is written $\nu(J)$, and the operation of expectation with respect to $\nu$ is denoted by $\bf E$ or $[\cdots ]_\nu$. 

While some central calculations can be carried out more generally, 
we mainly assume that the joint distribution of all the bonds is homogeneous, that is, invariant under some symmetry 
group that permutes the sites at which spins are located. For the finite-range cases, we assume that the sites $i$ label 
positions ${\bf x}_i$ in the lattice ${\bf Z}^d$, embedded in $d$-dimensional Euclidean space, and 
use the induced Euclidean metric to define distances on ${\bf Z}^d$, with the distance between nearest neighbors 
being $1$. Then we assume the distributions of the bonds are invariant under the group ${\bf Z}^d$ of translations.
We can identify subsets $\Lambda$ with sets of lattice sites, and use finite such portions $\Lambda$ of this set-up 
to define finite-size systems, that is with free boundary conditions; periodic boundary conditions, which ensure 
translation-invariance in a finite-size system, or other boundary conditions, can be handled with only minor modifications. 
All the general statements and bounds obtained below are independent of the boundary conditions used. In the basic 
EA model, the set $\cal E$ is the set of nearest-neighbor pairs, with the same value of ${\rm Var}\,J_{ij}=1$, say, for 
each such pair. Another interesting model is the power-law model \cite{kas,ks} (mentioned in the one-dimensional case 
in the introduction), in which $\cal E$ is the set of all pairs, and ${\rm Var}\,J_{ij}=|{\bf x}_i-{\bf x}_j|^{-2d\sigma}$. 

For infinite-range models, the basic example is the Sherrington-Kirkpatrick (SK) model \cite{sk}, in which for 
$i\in\Lambda=\{0,\ldots,N-1\}$, the 
joint distribution of bonds is invariant under the symmetric group $S_N$ that acts by permuting the sites, and 
the variance is ${\rm Var}\,J_{ij}=1/N$ for all $\{i,j\}$.

Both types of model can be extended to $p$-spin interactions for $p\geq 1$. For these, we use notation 
$X$, $Y$, \ldots for arbitrary finite subsets of $\Lambda$, and also define, for Ising spins, $s_X=\prod_{i\in X}s_i$. 
Then a $p$-spin Hamiltonian in system $\Lambda$ has the form \cite{derrida}
\be 
H=H(s)=-\!\!\!\!\sum_{X\subseteq \Lambda:|X|=p} J_Xs_X,
\ee
where the sum is over distinct subsets containing $p$ sites; this is called a $p$-spin model (for the specified $p$). 
Thus the EA Hamiltonian with $h=0$ is a $p=2$ model, and 
$p=1$ terms are random magnetic fields. Models that contain terms for more than one, or possibly all finite, $p$, 
\be 
H=-\!\!\!\!\sum_{X\in{\cal X}(\Lambda)} J_Xs_X,
\ee
where ${\cal X}(\Lambda)$ is the set of all finite subsets of $\Lambda$ [we set ${\cal X}({\bf Z}^d)={\cal X}$],
are called mixed $p$-spin models. The 
$X=\emptyset$ term cancels in the Gibbs weights, and so it can be set to zero; also, we write $J=(J_X)_{X\in{\cal X}(\Lambda)}$
and  $J|_{{\cal X}(\Lambda')}=(J_X)_{X\in{\cal X}(\Lambda')}$ for $\Lambda'\subseteq\Lambda$. 
For $\Lambda$ finite, the mixed $p$-spin Hamiltonians include 
all possible Ising Hamiltonians, because $\{s_X:X\subseteq \Lambda\}$ forms a complete set of 
functions of the spins, which is orthonormal with respect to the inner product defined by summation over all $s\in 
S_\Lambda$ with weight $2^{-|\Lambda|}$; from an arbitrary function $H(s)$, its expression as 
$-\sum_{X\subset \Lambda}J_X s_X$ (the Fourier-Walsh expansion) can be obtained by using orthonormality.
For mixed $p$-spin SG models, we again require the joint distributions of $J$ to be homogeneous, and 
there are both finite- and infinite-range versions; in the infinite-range SK-type versions, for $|X|=p$, 
the variance scales as ${\rm Var}\,J_X \propto N^{-(p-1)}$ \cite{derrida}. If desired, here (and also for the $m$-vector
models that follow) we can allow the $p=1$ term to 
have nonzero mean (a uniform magnetic field), with no effect on any of the later results.

For vector spins, we consider $m$-component unit vectors $s_i$ (which we also call $m$-vector spins) for each $i\in\Lambda$, 
and the standard inner product 
on ${\bf R}^m$, with $2$-spin interaction terms $-J_{ij}s_i\cdot s_j$ that possess $O(m)$ symmetry. The space of 
spin configurations is now $s\in S_\Lambda=({\bf S}^{m-1})^\Lambda$, where ${\bf S}^{m-1}$ is the unit sphere 
in ${\bf R}^m$. Summation over the Ising variables is now replaced by integration over $S_\Lambda$, using as measure 
the product of uniform [i.e.\ $O(m)$-invariant] measures which are the same on each ${\bf S}^{m-1}$ factor.
More generally, we could consider $m$-vector mixed $p$-site models, and require the interactions to be $O(m)$ 
invariant, at least for $p>1$; to be $O(m)$ invariant, a term for a set $X$ of $p$ sites must be a product of $2$-spin 
factors $s_i\cdot s_j$ ($i\neq j$), where for $m>1$, sites $i$,$j$ are allowed to appear more than once in the product. 
Thus for given $p>1$ and $m>1$ there is more than one way (in fact, a countable infinity of ways) to construct distinct 
such terms for each $X$ (with $|X|=p$), and so the terms in $H$ would not be indexed solely by $X$. $m>1$-vector 
spin models with anisotropic interactions would be similar. For all these cases, in general the $p\geq 1$ terms would have to be 
indexed using pairs $(X,x)$ in place of $X$, where $X$ is again a set of sites, and $x$ labels distinct interaction types 
for each $X$; then the random bonds would become $J_{(X,x)}$, the products $s_X$ become $s_{(X,x)}$ [constructed 
to obey $|s_{(X,x)}|\leq 1$ for all $(X,x)$] and sums would be taken over $(X,x)$ (we leave the trivial details of such changes 
to the following results to the reader). [We note that for the Fourier-Walsh expansion, for general $m>1$, 
for each $i$ the two basis functions, $1$ and $s_i$, in the Ising case ($m=1$) must be replaced by an infinite orthonormal 
set, which could be taken to be constructed from products of components of $s_i$, i.e.\ the traceless symmetric tensors on 
${\bf S}^{m-1}$, yielding infinite sums even for finite $\Lambda$.] 

The case of general discrete-spin models, such as Potts spins, is similar to the 
$m$-vector models. Such a spin variable can be represented as a discrete 
set of vectors in ${\bf S}^{m-1}$ for some $m$, and so will not be discussed further here.  
In every case, we assume that the underlying measure for sums or integrals over $s_i$ for each $i$ is uniform, 
meaning that it is invariant under a transitive action of a compact group of (measurable) symmetries, namely the group 
$O(m)$ for $m$-vector models, and the permutation group $S_p$ for $p$-state discrete spin models such as $p$-state 
Potts models. (This property will be used in Appendix \ref{app:gibbs}). Note that this statement holds regardless of whether 
the Hamiltonian is invariant under the group, though it could be. Clearly this general setup can be extended to spins that 
take values in other spaces, $S_0$ say, with a transitive action of a compact group as well. For discrete cases, 
the existence of a uniform measure that can be normalized as a probability distribution requires that $S_0$ be finite. 
For our purposes, the Ising ($m=1$) and $m$-vector ($m>1$) models provide sufficiently representative examples. 

Next we will extend the class of disorder distributions beyond Gaussians (readers can continue to think solely of Gaussians
if they prefer). We still require that the $J_X$ [or $J_{(X,x)}$] are centered (except possibly for $p=1$), have finite variance, 
are independent, and have a homogeneous joint distribution. It will be convenient to assume also that the marginal distribution of 
$J_X/\sqrt{{\rm Var}\,J_X}$ [$(J_X-{\bf E}J_X)/\sqrt{{\rm Var}\,J_X}$ for $p=1$] is the same for 
all $X$ [or all $(X,x)$], though this will be rarely used, and can easily be relaxed.  
A class of distributions that we could use is that of sub-exponential random variables, for which the probability density (relative 
to Lebesgue measure on the line) decays like the exponential function in (a constant times) $|J_X|$ at large $|J_X|$, or faster 
(see e.g.\ Ref.\ \cite{vershynin_book}). This condition ensures that the absolute moments of all orders are finite. We would 
then require that
\be
|{\bf E}J_X^n| \leq n! k^n ({\rm Var}\,J_X)^{n/2}
\label{subexp}
\ee
for all integers $n>0$ and all $X$, where $k$ is a constant, as in Refs.\ \cite{ks,vEvH_83}. These conditions imply that
${\bf E}e^{tJ_X}$ converges, and so is an analytic function of $t$, for sufficiently small $t$ [see Ref.\ \cite{vEvH_83}, 
eq.\ (2.13)], which then implies that $J_X$ is sub-exponential \cite{vershynin_book}. In practice, it 
will {\em not} be necessary to use the sub-exponential assumption, or conditions (\ref{subexp}), in this paper; similarly to
Refs.\ \cite{zegarlinski_91,cg_book}, the assumptions of independence, ${\bf E}J_X=0$ for $p>1$, ${\rm Var}\,J_X<\infty$,
homogeneity, and the same distribution of $J_X/\sqrt{{\rm Var}\,J_X}$ for all $X$ (modified for $p=1$ as above), will be 
the only general conditions imposed on the distributions. (In Sec.\ \ref{subsubothdis}, we will drop most of these conditions,
except for independence and translation invariance, and impose a weaker condition in place of the others.)

Now we complete the definitions of the finite- and infinite-range SG models. (In fact, for vector spins it is convenient to impose two 
additional restrictions, for which see the following Section \ref{subsubgibbs}.) In both cases, we 
would like to have a non-trivial thermodynamic limit as $|\Lambda|$ (or $N$) $\to\infty$ through a suitable sequence of models, 
for both the usual thermodynamic functions, such as the free energy, and also for the state, or distribution 
functions for the spins (equivalently, for their correlation functions). (For results for thermodynamics, see especially 
Refs.\ \cite{ks,vEvH_83,zegarlinski_91,gt,cg_book}, Ref.\ \cite{cg_book} for a review, and also Appendix \ref{app:energy}; 
for correlation functions and states in finite-range models at high temperature, see Refs. \cite{fz1,zegarlinski_87}.) We will show that
a sufficient condition (in addition to those of the preceding paragraph) for these to exist is
\be
\lim_{|\Lambda|\to\infty} \sum_{p\geq1}\sum_{X\subseteq \Lambda:i\in X,|X|=p}{\rm Var}\,J_{X} <\infty
\label{cvgce_cond}
\ee
(which is independent of $i$), which we call the convergence condition. (Here for Ising spins the $p=1$ term is irrelevant 
to the convergence of the sum, and can be dropped, as its variance is finite. Technical aspects of the definition and properties 
of such sequences of partial sums are given in Appendix \ref{app:gibbs}.)  This condition is clearly equivalent to finiteness 
of the sums for each $p$, together with convergence of the sum over $p$. For the finite-range translation-invariant models, 
we prove in Appendix \ref{app:gibbs} that non-trivial limits for the states exist (in a sense to be 
explained) at all nonzero temperatures and are Gibbs states (see Sec.\ \ref{subsubgibbs} below) when the convergence 
condition holds (a different approach that leads to a similar result but only at sufficiently high temperature can be found 
in Ref.\ \cite{zegarlinski_87}). We note that the sufficient condition for existence of a limit for thermodynamics involves 
convergence of a similar but (when more than one value of $p>1$ occurs in the sum) different sum \cite{zegarlinski_91,cg_book}; 
the above condition implies that condition also (see Appendix \ref{app:energy}). 
Thus for the finite-range models, ${\rm Var}\,J_X$ can be taken independent of $|\Lambda|$, but must fall off as the distances 
between the sites in $X$ increase, such that the sum converges. 

The models we call infinite range necessarily involve instead 
${\rm Var}\,J_X$s that depend explicitly on $N=|\Lambda|$ (that is, they tend to zero as $N\to\infty$), as already specified 
for SK-type $p$-spin models, such that 
they obey the same convergence condition (\ref{cvgce_cond}), in which all terms contribute in the limit. Then again the thermodynamic 
limit of the free energy exists under a similar, though different, condition which is implied by this one, plus a convexity 
condition \cite{gt,cg_book}. The $d=1$, $p=2$ power-law model defines a finite-range model for 
$\sigma>1/2$, but for $0\leq \sigma< 1/2$ ${\rm Var}\,J_{ij}$ must scale as $|\Lambda|^{2\sigma-1}$ (or 
$1/\ln |\Lambda|$ for $\sigma=1/2$) in order to produce correct behavior of the limit in this case; thus for those 
values of $\sigma$ this model is infinite-range, but not strictly of SK type except when $\sigma=0$. Finally, 
finite-range models, such as the EA model, in which for each $i$ there are nonzero (and 
$|\Lambda|$-independent) terms in $H$ for only a finite number of $X$ with $i\in X$ will be called strictly short 
range [some might use the term ``finite'' (i.e.\ bounded) range for those instead, but we will not]. In the following Section 
we will also subdivide finite-range models into long range and short range, with strictly
short range a subset of the short-range models.

\section{Complexity bounds for finite-range spin glasses}
\label{sec:compl}

Now we focus on finite-range SGs. In Section \ref{subgibbs}, we discuss Gibbs states and metastates in SGs from a 
somewhat informal point of view, including the cases of $m$-vector spins or long-range interactions. In Section \ref{subinfo}, 
we explain some basic definitions and results from information theory that will be used in the paper. Finally, in Section \ref{subcomp}, 
we combine the concepts to define complexity as mutual information, introduce basic results, and derive the upper bounds advertised 
in the Introduction. The basic methods used for the bounds in finite size are quite elementary, and are similar to methods for bounding
surface free energies (see e.g.\ Ref.\ \cite{cg_book}). The final (and essential) extension to results for infinite size uses some results 
from the Appendices, especially in the long-range case. Section \ref{subcomp} ends with the extension to more general disorder 
distributions, and further physical discussion.

\subsection{Gibbs states, pure states, and metastates}
\label{subgibbs}
\subsubsection{Gibbs states and short-range models}
\label{subsubgibbs}

In a finite-size system, given a Hamiltonian $H=H(s)$, the Gibbs distribution on spin configurations $s=(s_i)_{i\in\Lambda}$
($\Lambda$ finite) at temperature $T$ is given by the well-known formula
\be
p_H(s)=e^{-H/T}/\sum_s e^{-H/T}.
\label{gibbsfin}
\ee
(In general, our convention is to treat the spins as discrete variables, with evident generalizations to continuous variables such as 
$m$-vector spins.) This definition does not work when the set $\Lambda$ becomes infinite, due to the infinite sum in $H$. Instead, 
the preliminary definition \cite{georgii_book,simon_book,bovier_book,liggett_book} of an infinite-size Gibbs state is as a 
probability distribution (also called a state) $\Gamma(s)$ on spin configurations 
$s=(s_i)_{i\in{\bf Z}^d}$ such that for any finite subset $\Lambda$ of ${\bf Z}^d$, the conditional probability distribution 
for spins $s|_\Lambda$ given $s|_{\Lambda^c}$ (for any $\Lambda$, $\Lambda^c={\bf Z}^d-\Lambda$ is its complement)
is
\be
\Gamma(s|_\Lambda\mid s|_{\Lambda^c})=p_{H'}(s|_\Lambda),
\label{gibbsdef}
\ee
where 
\be
H'=-\sum_{X\in {\cal X}:\Lambda \cap X\neq\emptyset}J_X s_X,
\label{H'def}
\ee
so $H'$ [or $H'_\Lambda(s)$)] contains all interactions entirely within $\Lambda$, as well as all interaction terms involving
at least one spin in $\Lambda$ and at least one of the fixed spins in $\Lambda^c$. (Note that in our notation, $p_{H'}(s|_{\Lambda})$ 
depends implicitly on $s|_{\Lambda^c}$ through $H'$, while the sum in eq.\ (\ref{gibbsfin}) is here over $s|_\Lambda$ only.) 
As ${\bf Z}^d$ is infinite, the definition
of conditional probability needs care; see Ref.\ \cite{chung_book,breiman_book}. In general, $\Gamma(s)$ may not be
uniquely determined by this definition, and that may lead to interesting physics, namely existence of many pure states. 
We note that we will write the (conditional) 
probabilities for a specific vector $s$ value as we did here \cite{chung_book,breiman_book}, but $\Gamma$ is really a 
measure, and we will also write, for example, $\Gamma(A)$ for the thermal probability of a {\em set $A$} of values of $s$. 
We write $\rm E$ or $\langle\cdots\rangle$ for the operation of (possibly conditional) expectation over $s$ in a finite- 
or infinite-size Gibbs state. A state $\Gamma$ may be a Gibbs state for a specific $J$, but we will not show that explicitly 
in the notation. 

The definition makes sense in the strictly short-range Ising case, because the sum over $X$ is finite, and further the sum is finite for 
almost all $J$ (i.e.\ with probability $1$ with respect to $\nu$). For finite-range models more generally, there is an obvious 
convergence issue regarding the sum over $X$ in $H'$. In the remainder of this 
paragraph we restrict ourselves to Ising spins, with simple modifications for more general discrete spins with $S_0$ finite; we return 
to vector spins afterwards.  The definition continues to make sense if the limits of partial sums (for 
fixed $\Lambda$),
\be
\lim_{|\Lambda'|\to\infty}\sum_{X\subseteq \Lambda':\Lambda \cap X\neq\emptyset}J_X s_X,
\label{energysum}
\ee
exist and are finite for all $s$ and all finite $\Lambda$ (again, see Appendix \ref{app:gibbs} for the precise definition of the limit). 
For fixed $\Lambda$ and $J$, convergence for all $s$ is guaranteed if the sum converges absolutely for some $s$. In general, 
the corresponding partial sums $\sum_{X\subseteq \Lambda':X\cap \Lambda \neq\emptyset}|J_X|$ have a finite limit as 
$|\Lambda'|\to\infty$ for $\nu$-almost all $J$ if the sum of expectations 
\be
\sum_{X\subseteq \Lambda':X \cap\Lambda \neq\emptyset}{\bf E}|J_X|
\ee converges. (This follows 
from the monotone convergence theorem, i.e.\  ${\bf E}f_n(J) \to{\bf E} f(J)$ for an increasing sequence of non-negative functions 
$0\leq f_n\nearrow f$ as $n\to\infty$ for almost all $J$ \cite{chung_book,breiman_book}, applied to the partial sums, 
together with the fact that, for $f\geq 0$, ${\bf E} f<\infty$ implies $f<\infty$ for almost all $J$.) Under our conditions on
the distribution of $J_X$, that occurs if 
\be
\lim_{|\Lambda'|\to\infty} \sum_{X\subseteq \Lambda':\Lambda\cap X\neq\emptyset}\sqrt{{\rm Var}\,J_{X}}
\ee
is finite. This condition holds for all finite $\Lambda$ if and only if 
\be
\lim_{|\Lambda'|\to\infty} \sum_{p>1} \sum_{X\subseteq \Lambda':i\in X,|X|=p}
\sqrt{{\rm Var}\,J_{X}}<\infty;
\label{sh_cvgce_cond}
\ee 
this is clearly equivalent to the sums for each $p>1$ converging, together with
convergence of the sum over $p$, and also implies that condition (\ref{cvgce_cond}) holds. If a finite-range model 
satisfies this condition, then we call it short range; otherwise, it is long range \cite{vEvH_83}. (The interactions in 
short-range models are also called regular or absolutely summable \cite{bovier_book}; those in the long-range models 
are also called square summable.) In the long-range case, the definition of a Gibbs state 
has to be modified somewhat, as given in Section 2 of Ref.\ \cite{gns}, and requires further discussion; see 
Appendix \ref{app:gibbs}, and also Ref.\ \cite{zegarlinski_87} for a related treatment. 

For $m$-vector spins, or other continuous spins, the sum of $|J_{(X,x)}|$ over $(X,x)$ could diverge even when the 
system size or $\Lambda'$ is finite. To avoid this possibility, we consider only finite-range models 
that obey, in addition to the previous general conditions and condition (\ref{cvgce_cond}), both (i)
\be
\sum_{p>1} \sum_{(X,x):X\subseteq \Lambda',i\in X,|X|=p}
\sqrt{{\rm Var}\,J_{(X,x)}}<\infty
\ee
for all finite $\Lambda'$, and (ii), for the $p=1$ terms,
\be
\sum_{(X,x):X=\{i\}}{\bf E}|J_X|
\ee 
converges (for all $i$).
Then as before, when the limit $\Lambda'\to\infty$ of the sum in (i) is finite, we define that to be the
short-range case; if the  $\Lambda'\to\infty$ limit is infinite, but the corresponding sum of variances converges, we define that 
to be the long-range case. In this way the names remain apt, 
the interactions among any given finite set $\Lambda'$ of spins $s|_{\Lambda'}$ (and also the $p=1$ single-site terms) 
are fairly well behaved, and only minimal changes are required for $m$-vector spins in the proofs of the results.
We impose the same condition for infinite-range continuous spin models. 
The additional stipulations for continuous spins will usually be left implicit from here on.
More general models than these are outside the scope of this paper.

It may be helpful here to consider the example of the $2$-spin power-law models in dimension $d$. Then we
have the well-known results that the models are finite range when $\sigma>1/2$, with a well-defined limit for the 
thermodynamic properties, and short range when $\sigma>1$. We emphasize that the distinction that we use in this
paper between short range and long range models is important technically, but its physical significance for general $d$ 
is less clear in general. A change in the behavior of our complexity bounds at a value of $\sigma$ that is different from $1$
for $d>1$ will follow from the later results. For the $d=1$ power-law $2$-spin 
model \cite{kas}, the critical value of $\sigma$ at the short-long boundary is also the value above which 
the transition at $T>0$ disappears; this is a consequence of being in one dimension. For analogous physical 
discussion of higher $d$, see Refs.\ \cite{fh,bmy}. 

\subsubsection{Pure states}

Once Gibbs states for a given Hamiltonian $H$ (thus, for given $J$) are at hand, we notice that a convex combination 
(or mixture) of distinct Gibbs states for the same $H$ is again a Gibbs state; in general, a convex combination could 
involve an average taken using a probability measure on Gibbs states. A Gibbs state that cannot be expressed as such 
a combination of other Gibbs states is called an extremal or pure state. Any Gibbs state can be decomposed uniquely 
into a mixture of pure states \cite{georgii_book,simon_book,bovier_book,liggett_book}, in the form
\be
\Gamma(s)=\sum_\alpha w_\alpha (\Gamma)\Gamma_\alpha(s),
\label{puredec}
\ee 
where $\Gamma_\alpha$ are pure states, and the weights $w_\alpha\geq 0$ sum to $1$. (Again, in practice the decomposition 
might be continuous and require an integral over $\alpha$ using a measure $w(\alpha)$ in place of the discrete weights
$w_\alpha$, but we will usually not show this explicitly.) The complexity of this decomposition in a SG is one of the 
main topics investigated here. 

\subsubsection{Metastates and metastate-average state}

In general, there may be many Gibbs states for a given $H$ (especially at low temperature), and their 
relevance physically may not be obvious.
To obtain states of physical relevance, we can use some sort of limit of finite-size systems; the metastate concept 
provides such a construction \cite{ns96b,ns_rev,aw}. Here we will describe the operational content of the construction 
\cite{read14}, leaving more rigorous discussion to Appendix \ref{app:gibbs}. The existence of a unique Gibbs state 
that is the limit of finite-size states is far from clear (again, particularly at low temperature), and the metastate provides 
instead a probability distribution on Gibbs states, for given $J$. To describe the pioneering Aizenman-Wehr (AW) construction 
\cite{aw}, we first introduce hypercubes such as $\Lambda_W$ of side $W$, which is centered at the origin in 
${\bf Z}^d$ ($W$ is odd), and similar hypercubes $\Lambda_R$, $\Lambda_L$ for odd $R$, $L$, where $W<R<L$. 
We start with a finite-size system in $\Lambda=\Lambda_L$ to produce a Gibbs state $\Gamma_L=p_H(s|_{\Lambda_L})$, where
\be
H=-\sum_{X:X\subseteq \Lambda_L}J_Xs_X
\ee 
(one can use small variations on this choice of $H$ at the boundary of $\Lambda_L$ to obtain other boundary conditions, including 
the case of periodic boundary conditions).
With this we can calculate thermal averages of functions of the spins $s|_{\Lambda_W}$ in $\Lambda_W$ for the given $J$,
or functions of such averages; these depend on $J$ in general. An average over the finite-size analog of the metastate 
is now a disorder average (expectation) using $\nu(J|_{{\cal X}(\Lambda_L)-{\cal X}(\Lambda_R)})$, where 
the disorder is (at least partially) outside $\Lambda_R$ (as the distribution is a product, it can be viewed as conditioned on 
$J|_{{\cal X}(\Lambda_R)})$. Finally, one can average (a function of)
the metastate average using $\nu(J|_{{\cal X}(\Lambda_R)})$. We then take limits as $L\to\infty$, then $R\to\infty$,
if these exist. In the limit, the final $\nu(J|_{\Lambda_R})$ is viewed (and written) as disorder average using $\nu(J)$,
while the average over the original outer region is denoted as metastate average using the measure $\kappa$, which is a probability
distribution $\kappa(\Gamma)$ on Gibbs states $\Gamma$ for given $J$ (i.e.\ those $J$ in what was formerly the inner 
region $\Lambda_R$ before $R\to\infty$). 
Likewise, the Gibbs states obtained, written $\Gamma(s)$, are distributions on spin configurations $s$, obtained as the limit as
$W\to\infty$ subsequent to the prior limits. An AW metastate contains information about the extent to which the Gibbs state 
$\Gamma$ (at given $J$) depends on the disorder asymptotically far away.
The technical question of existence of the limits can be handled by using subsequences, at the cost of possible non-uniqueness of
the metastate, and the fact that the infinite-size states are Gibbs states has to be proved. $\kappa$ or $\kappa_J$ 
depends on the bonds $J$ through the Gibbs states $\Gamma$, but we will leave that implicit. As in the case of the pure states 
in the pure state decomposition, we usually treat $\Gamma$ as a discrete variable, though the distribution $\kappa$ may in fact 
be continuous.

In the NS metastate construction \cite{ns96b,ns_rev}, the use of the disorder average over 
$J|_{{\cal X}(\Lambda_L)-{\cal X}(\Lambda_R)}$ is replaced by an 
empirical average over a range of sizes, say $R=L_0<L_1< \cdots <L_K=L$. Again taking $L$, $R$, and $K$ to infinity 
(possibly using a subsequence) produces the metastate average. With a suitable choice of such a subsequence of sizes, 
a NS metastate can be shown to be the same as a corresponding AW metastate. The NS 
metastate contains information about the extent to which, asymptotically, the Gibbs state $\Gamma$ (at given $J$) 
varies with system size in the finite sizes from which it was constructed.  It too will be denoted $\kappa$, as the 
following results apply equally to either metastate. In either construction, the question of uniqueness of the metastate 
in a SG remains open. 

Finally, there is one additional construction that will be used. In either metastate construction, the Gibbs states can be 
averaged using the metastate, to produce the metastate-average state (MAS) or barycenter $\rho(s)$ \cite{aw,ns96b,ns_rev}, 
which is itself a Gibbs state (though a rather special one), and it still depends on the sample of $J$. That is, 
$\rho(s)=[\Gamma(s)]_\kappa$, where the square bracket $[\cdots ]_\kappa$ denotes metastate average.
Then the MAS has pure state decomposition
\bea
\rho(s)&=&\sum_\Gamma\kappa(\Gamma)\sum_\alpha w_\alpha (\Gamma)\Gamma_\alpha(s)\\
&=&\sum_\alpha \mu_\alpha \Gamma_\alpha(s),
\eea
where the weights are $\mu_\alpha=\sum_\Gamma \kappa(\Gamma) w_\alpha(\Gamma)$.
The MAS $\rho$ can also be obtained more directly from finite size, by taking the average of the finite-size Gibbs state 
$\Gamma_L$ over the disorder in the outer region using $\nu(J|_{{\cal X}(\Lambda_L)-{\cal X}(\Lambda_R)})$, and then 
taking limits along the same subsequences as before. 

\subsection{Information theory concepts}
\label{subinfo}
\subsubsection{Mutual information and relative entropy}
\label{subsubinfo}

Next we define the concepts from information theory that we will use (see e.g.\ Ref.\ \cite{ct_book}). First, we 
have the mutual information of two random variables, possibly conditioned on a third. Suppose that
$\cal A$ ($\cal B$, $\cal C$) is a random variable, which takes values that we generically 
call $a$ (respectively $b$, $c$). To lighten notation we will treat the values as discrete, and denote
the joint probabilities $p(a,b,c)$; notation such as $p(a,b)$ will mean the corresponding marginal
probabilities for $a$, $b$ alone---that is, $p(a,b)=\sum_c p(a,b,c)$, and so on---and the conditional 
probabilities are given implicitly by, for example, $p(a,c)=p(a|c)p(c)$. Then the mutual
information between $\cal A$ and $\cal B$, given $\cal C$, is defined as
\be
I({\cal A};{\cal B}|{\cal C})=\sum_{a,b,c}p(a,b,c)\ln\frac{p(a,b|c)}{p(a|c)p(b|c)}.
\ee
(The convention $0\ln 0=0$ is used.)
The unconditional mutual information $I({\cal A};{\cal B})$ is given by the same formula with $c$ 
and conditioning on $c$ deleted everywhere, or one may consider the case in which $\cal C$ is a constant; 
the same applies to subsequent definitions also. Our 
random variables such as $\cal A$ do not have to be real-valued; they could 
take other values, including vectors. Consequently, any of them could be interpreted as two or more 
random variables, leading to the definitions of $I({\cal A}_1,\ldots,{\cal A}_k;{\cal B}_1,\ldots,
{\cal B}_n|{\cal C}_1,\ldots,{\cal C}_m)$ similarly. The (conditional) mutual 
information can be related to the (conditional) Shannon entropy,
\be
S({\cal A}|{\cal C})= - \sum_{a,c}p(a,c)\ln p(a|c),
\ee
by 
\be
I({\cal A};{\cal B}|{\cal C})=S({\cal A}|{\cal C})-S({\cal A}|{\cal B},{\cal C}).
\ee

We also define the relative entropy (or Kullback-Leibler divergence) of two (conditional) probability distributions, which will
appear in calculations later. If $p(a,c)$, $q(a,c)$ are two distributions for the same random variables $\cal A$, 
$\cal C$, then the conditional relative entropy [i.e.\ of $p(a|c)$ relative to $q(a|c)$, and using $p(c)$ implicitly] is defined as 
\cite{ct_book}
\be
D\left[p(a|c)||q(a|c)\right]=\sum_{a,c}p(a,c)\ln\frac{p(a|c)}{q(a|c)}.
\ee
(Here the left-hand side is a functional of the probability distributions, not a function of $a$, $b$, $c$.)
$I({\cal A};{\cal B}|{\cal C})$ can be obtained from $D$ by replacing $\cal A$ by $\cal A$, $\cal B$, and taking 
$q(a,b|c)=p(a|c)p(b|c)$, the  product distribution formed from the (conditional) marginals. 

While we will continue to use the notation as for the 
discrete case, we note here that for continuous variables, the probabilities such as $p(a)$ must be replaced by
the use of a probability measure $P$ say, where the probability that $a$ is in a set $A$ is $P(A)$ (here $A$ is an element
of a suitable $\sigma$-algebra of measurable sets; in general, in the main text we will not need to concern ourselves with 
such issues), and expectation over $a$ is expressed as the integral constructed from the measure $P$. 
In order to make use of definitions of $I$ and $D$ similar to the preceding ones for the discrete case, we can proceed 
in either of two equivalent ways. For the first, it may be that the probability $P(A)$ can, for all $A$, be expressed 
in terms of a density $p(a)$ (i.e.\ a measurable function of $a$) with respect to some reference measure $P_0$, such that 
$P(A)=\int_A p(a)P_0[da]$, where the integral is taken over the subset $A$ in the set of all $a$ values. If this can
also be done for another measure $Q$ with density $q$ relative to the same reference $P_0$, for variables $a$, $b$, and $c$, 
then the definition of $D$ takes the same form as above except that the sums are replaced with the corresponding integral with 
respect to $P_0$; a similar statement holds for $I$ \cite{ct_book}. We note that $I$ and $D$ are invariant under a change of 
integration variable(s) of the form $a\to a'(a)$, $b\to b'(b)$, $c\to c'(c)$ or under a change of reference measure $P_0$
[both types of change involve changing $p$ and $q$ by a (common) Jacobian factor so that the measures $P(A)$, $Q(A)$ 
of a set $A$ of $a$, $b$, $c$, \ldots are unchanged]. The alternative, more general, approach uses a partition of the space 
of $a$, $b$, $c$, \ldots, 
into a discrete set of disjoint subsets, for which the definitions above in the discrete case can be used; $D$ or $I$ can 
then be defined as the limit of such quantities as the partition becomes arbitrarily fine \cite{ct_book,pinsker_book}. Either 
approach gives well-defined results for those quantities. 

As an aside, we note that for the Shannon entropy and conditional entropy in the case of a continuous distribution, the 
two corresponding approaches are not equivalent, and nor are they well defined. The (conditional) entropy of a partition diverges as 
the partition becomes arbitrarily fine, while the (conditional) differential entropy using the density relative to $P_0$ 
differs from the former by subtraction of a term that diverges in the limit \cite{ct_book}, and also is not independent of $P_0$.
(These effects cancel in $I$ and $D$.) This leads to additional difficulties for Palmer's definition of complexity as entropy 
\cite{palmer}, if the distribution $w$ (or $\mu$) of pure states is continuous.

An application of Jensen's inequality 
shows that $I$ and $D$ are both non-negative, including in the continuous case; $D$ is zero if and only if 
$p(a|c)=q(a|c)$ for all $a$ (for all $c$ that have nonzero probability), and hence $I$ is zero if and only if $\cal A$ and 
$\cal B$ are (conditionally) independent, that is $p(a,b|c)=p(a|c)p(b|c)$ (for all $c$ that have nonzero probability) \cite{ct_book}.
Thus $D$ is a (non-symmetric) non-negative measure of the distance of the distribution $p(a|c)$ from $q(a|c)$, averaged over $c$,
and $I$ is a non-negative measure of how far $p(a,b|c)$ is from the product of (conditional) distributions. 
Hence mutual information is a measure of how much information one gains about the value of $\cal B$ from an observation 
of the value of $\cal A$ (or {\it vice versa}), conditioned on the value of $\cal C$, with again the final $c$ average. $I$ is also 
a good way to measure the correlation between two subsystems, as it is independent of a choice of some representative
random variable for each subsystem, as must be made in a conventional correlation function.
There is a remaining issue of whether $I$ and $D$ are {\em finite}, 
i.e.\ non-diverging, to which we return later. 

\subsubsection{Chain rule and Markov property}
\label{subsubchain}

An elementary calculation from the definitions leads to the chain rule formula \cite{ct_book},
\be
I({\cal A};{\cal B}_1,{\cal B}_2|{\cal C})=I({\cal A};{\cal B}_1|{\cal C})+I({\cal A};{\cal B}_2|{\cal B}_1,{\cal C}),
\ee
and also the same with ${\cal B}_1$, ${\cal B}_2$ interchanged. It can be extended to various multiterm formulas when, 
for example, ${\cal B}_2$ is replaced by ${\cal B}_2$, \ldots, ${\cal B}_n$, by iterating the use of the 
above formula. [There is a similar version of the chain rule for (conditional) relative entropy.] Because the terms on the 
right-hand side are non-negative, the chain rule for mutual information immediately implies various inequalities, such as
\bea
I({\cal A};{\cal B}_1,{\cal B}_2|{\cal C}) &\geq& I({\cal A};{\cal B}_1|{\cal C}) ,\\
I({\cal A};{\cal B}_1,{\cal B}_2|{\cal C}) &\geq& I({\cal A};{\cal B}_1|{\cal B}_2,{\cal C}) .
\eea

An important application, which will be used repeatedly in the following, arises when there are three 
random variables, 
say $\cal A$, $\cal B$, $\cal C$, that constitute a Markov chain, say in the form ${\cal C}\to{\cal B}\to{\cal A}$ 
\cite{ct_book}. By definition, this means that $\cal A$ and $\cal C$ 
are independent when the value $b$ of $\cal B$ is given, in other words conditionally on $\cal B$: $p(a,c|b)=p(a|b)p(c|b)$. 
(As the definition is symmetric, the arrows could all be reversed, or replaced by double-headed arrows.) Equivalently,
$p(a|b,c)=p(a|b)$ for all $c$, which is another standard definition of the Markov property. Then the chain rule can be
applied in two ways: in general,
\bea
I({\cal A};{\cal B},{\cal C})&=&I({\cal A};{\cal C})+I({\cal A};{\cal B}|{\cal C})\\
&=&I({\cal A};{\cal B})+I({\cal A};{\cal C}|{\cal B}).
\eea
For the Markov chain ${\cal C}\to{\cal B}\to{\cal A}$, in the last line $I({\cal A};{\cal C}|{\cal B})=0$, so we have
\be
I({\cal A};{\cal B})=I({\cal A};{\cal C})+I({\cal A};{\cal B}|{\cal C}).
\ee
As each term is non-negative, this yields inequalities
\bea
I({\cal A};{\cal C}) &\leq& I({\cal A};{\cal B}),\\
I({\cal A};{\cal B}|{\cal C}) &\leq& I({\cal A};{\cal B}).
\eea
The first of these is known as the data-processing inequality \cite{ct_book} or as the pipeline inequality, the latter
because the information that can be transmitted to $\cal A$ from $\cal C$ cannot be greater than that from $\cal B$,
because it must ``come through'' $\cal B$. The second says that, for the Markov chain, conditioning on $\cal C$ 
reduces (or cannot increase) the mutual information between $\cal A$ and $\cal B$. 

\subsubsection{Infinite mutual information}
\label{subsubinfinite}

Now we return to the question of whether the mutual information or the relative entropy can be infinite. 
In formal treatments, it is common to define the (conditional) relative entropy only when $p(a|c)$ is absolutely continuous
with respect to $q(a|c)$. This condition requires that if $q(a|c)$ is zero for some $a$ and $c$, then $p(a|c)$ is zero also. 
If this condition does not hold for some $c$ with $p(c)>0$, then one can see that $D$ will be infinite, and that is how 
we will regard it, rather than imposing the condition of absolute continuity on $p$, $q$ as part of the definition. (Here 
we have in mind the discrete case; we discuss the continuous case afterwards.) For the mutual information, 
$p(a|c)=\sum_b p(a,b|c)$, and so, when $a$ and $b$ are discrete, absolute continuity always holds relative to the product 
distribution $p(a|c)p(b|c)$. The relative entropy and mutual information could still diverge when the final sum is over an 
infinite set.

For the continuous case, absolute continuity means that if a set $A$ of $a$ values is given zero probability by $Q$, 
then $A$ is also given zero probability by $P$. When $P$ and $Q$ are represented by functions (densities) relative 
to $P_0$, this simply puts conditions on the behavior of $p$ at the zeroes of $q$. But the absolute continuity would 
certainly be violated if, for example, $Q$ were represented by a genuine density $q$ (an ordinary function of $a$) relative 
to $P_0$, but the so-called density $p$ for $P$ contained a $\delta$-function of some coordinate, in which case of course 
$p$ would not really be a function. Then $D$ would again be regarded as infinite. For the mutual information $I$, 
the previous argument (with the sum replaced by an integral) for absolute continuity of $p$ relative to the product distribution 
may now fail if $p(a,b|c)$ contains a $\delta$-function rather than being a function. The density for the product distribution, 
$p(a|c)p(b|c)$, may contain no $\delta$-function, even though $p(a,b|c)$ does, yielding infinite $I$, as we will show with 
a physical example in a moment. Before that, we state the general result, which is that, in all cases, if the relative entropy 
of $P$ relative to $Q$ is finite, then $P$ is absolutely continuous with respect to $Q$ \cite{pinsker_book}. The converse 
does not hold in general.

For an example, we consider two unit $m$-vector spins $s_1$, $s_2$, with $m>1$ and ferromagnetic Hamiltonian 
$H=-Js_1\cdot s_2$ ($J>0$), at temperature $T\geq0$. At $T>0$, the Gibbs distribution $p(s_1,s_2) 
\propto e^{-H(s_1,s_2)/T}$ is continuous, and so are the marginal distributions $p(s_i)$ ($i=1$, $2$), obtained 
by integrating out the other spin. At zero temperature, we have $s_1=s_2$ with probability one, which means that 
the joint distribution contains a $\delta$-function in the relative angle coordinates between the two spins on the 
unit sphere. By symmetry, the marginal distributions 
$p(s_i)$ are uniform on the unit sphere for all $T\geq0$. Hence at zero temperature the mutual information 
$I({\cal S}_1;{\cal S}_2)$ of the two spins is infinite. This is quite physical; the mutual information increases 
steadily as $T\to0$, due to the increasing degree of correlation between the two spins (it diverges as 
$I\sim (m-1)\ln\sqrt{J/T}$ as $T\to0$). At $T=0$, the infinite value of $I$ reflects the fact that knowledge
of one spin gives knowledge of the other with infinite precision. For the Ising case $m=1$ with spin-flip 
symmetry, the mutual information instead tends to $\ln 2$ as $T\to0$. 

In SGs, when we examine the mutual information between two sets of spins (as discussed in Sec.\ \ref{subcomp} below), 
for vector spins in the absence 
of magnetic fields our models may have global $O(m)$ rotation symmetry (for $m=1$, this reduces to ${\bf Z}_2$, the spin-flip
symmetry of some of our models). Then at $T=0$ for $m>1$ the joint distribution will include $\delta$-functions in the relative 
angle coordinates between one set and the other, though now these will be coordinates on $SO(m)$ rather than the unit sphere in 
${\bf R}^m$. Hence a similar divergence in the mutual information will occur. This is clear for any Gibbs state in finite size, and again 
is replaced by $\ln 2$ for $m=1$ in the presence of ${\bf Z}_2$ symmetry. 
For all $m\geq 1$, it should be possible to factor off this effect of symmetry to leave mutual information that tends to zero
as $T\to0$ in any given $O(m)$-invariant Gibbs state drawn from the metastate.
For the MAS, it may be that this is the only effect giving rise to such a divergence. That is, if the global rotation of the 
ground-state spin configuration were factored out, the remaining distributions might give finite mutual information. For SGs, 
it is a non-trivial question whether this will be true. 

\subsection{Complexity of Gibbs states and metastates}
\label{subcomp}
\subsubsection{Complexity as mutual information}
\label{subsubcomp}

Next we apply mutual information to study the complexity of Gibbs states and metastates. In Ref.\ \cite{hr}, 
H\"oller and the author introduced a definition of the complexity of a Gibbs state as the mutual information between 
the spin configuration and the pure states; it was motivated by the earlier idea of a difference of entropies \cite{vEvH}. 
A Gibbs state $\Gamma$ drawn from the metastate has a decomposition as in eq.\ (\ref{puredec}),
with weights $w_\alpha(\Gamma)$. As the pure states $\Gamma_\alpha$ are fixed for the given $J$, 
we can identify $\Gamma$ with the set of $w_\alpha(\Gamma)$. The probability of $\Gamma$ is $\kappa(\Gamma)$.
Now let $\cal S$, $\cal A$, $\cal G$ be random variables with values $s$, $\alpha$, $\Gamma$ respectively; 
then their joint distribution is 
\be
\kappa(\Gamma)w_\alpha(\Gamma)\Gamma_\alpha(s).
\ee
$\kappa(\Gamma)$ is the marginal
distribution for $\cal G$, while the conditional distribution for $({\cal A},{\cal S})$ given ${\cal G}=\Gamma$ is 
$w_\alpha(\Gamma)\Gamma_\alpha(s)$. 

Even without mentioning a metastate, for a given $\Gamma$ we can form the mutual information $I({\cal S};{\cal A})_\Gamma$ 
[using the joint distribution $w_\alpha(\Gamma)\Gamma_\alpha(s)$] which quantifies the amount 
of information obtained about which pure state $\cal A$ the system is in from an observation of $\cal S$, the spin configuration 
of the whole system, for the given $\Gamma$. If $S({\cal S})$ is the entropy of the distribution $\Gamma(s)$, and 
$S_\alpha({\cal S})$ is that of $\Gamma_\alpha(s)$, and we ignore the fact that both are infinite in infinite size, then 
the mutual information is $I({\cal S};{\cal A})_\Gamma=S({\cal S})-\sum_\alpha w_\alpha(\Gamma)S_\alpha({\cal S})$ 
\cite{hr}; cf.\ eq.\ (3.20) 
in Ref.\ \cite{vEvH}, where they use an additional assumption that any spin configuration $s$ determines a unique $\alpha$, 
and see HR for further discussion. If we average it over $\Gamma$ as well, using the metastate, then we can identify the result as 
$I({\cal S};{\cal A}|{\cal G})$; we call this the (metastate average of the) complexity of a (typical) Gibbs state. 
As the MAS $\rho$ is itself a Gibbs state, we can similarly define its complexity $I({\cal S};{\cal A})_\rho=I({\cal S};{\cal A})$, where 
the probability used is the marginal $\mu_\alpha\Gamma_\alpha(s)$ for $({\cal S},{\cal A})$ obtained by summing over 
$\Gamma$.  As either of these mutual 
informations may diverge, a more conservative strategy is to replace $\cal S$ in the definitions with
${\cal S}_\Lambda$ with values $s|_\Lambda=(s_i)_{i\in\Lambda}$. We can then examine how this quantity grows with the size of 
$\Lambda$. This strategy can be viewed as calculating $I$ using a partition, indexed by $s|_\Lambda$, of the space 
$S$, as discussed in Sec.\ \ref{subsubinfo}. This is what was actually used in HR; it can also be done in the following formulas of this subsection, and will be used below,
but we leave it implicit for now. 

In addition, we can also define the complexity of the metastate itself to be $I({\cal S};{\cal G})$, obtained by using the marginal 
$\kappa(\Gamma)\Gamma(s)$; this gives the amount of information about the Gibbs state 
$\cal G$ obtained from an observation of the spins $\cal S$, ignoring the decomposition into pure states. It vanishes for
a trivial metastate (i.e.\ again, one supported on a single Gibbs state for given $J$). For the AW metastate construction, 
it is an infinite-size limit of the mutual information between the 
spins $\cal S$ (alternatively, the spins in a finite window, ${\cal S}_\Lambda$) and the disorder in the outer region.

Summarizing these definitions, we introduce symbols for the complexity $K_\Gamma$ of a (typical) Gibbs state, $K_\kappa$ of
a metastate, and $K_\rho$ of a MAS:
\bea
K_\Gamma&=&I({\cal S};{\cal A}|{\cal G}),\label{compgam}\\
K_\kappa&=&I({\cal S};{\cal G}),\label{compkap}\\
K_\rho&=&I({\cal S};{\cal A}).
\label{comprho}
\eea
All of these depend implicitly on $J$; the first two depend on the metastate $\kappa$ from which a $\Gamma$ is drawn, 
while the last depends only on the MAS $\rho$. 
For the versions with a finite window $\Lambda$ for the spins ${\cal S}_\Lambda$, we write $K_\Gamma(\Lambda)$ and so on.
In the case of discrete spins and strictly short range interactions, and for $T\geq 0$, a bound proportional to the 
surface area of $\Lambda$, similar to that in HR \cite{hr}, applies to all three of these, including the complexity of the metastate.

An important observation is that in general the three random variables form a Markov chain, 
\be
{\cal G}\to{\cal A}\to{\cal S}.
\ee
This is because the conditional distribution for $\cal G$ and $\cal S$, given ${\cal A}=\alpha$, is 
\be
\frac{\kappa(\Gamma)w_\alpha(\Gamma)}{\sum_\Gamma\kappa(\Gamma)w_\alpha(\Gamma)}\Gamma_\alpha(s)
\ee
(using $\sum_s\Gamma_\alpha(s)=1$), which is a product of the marginals conditioned on ${\cal A}=\alpha$. 
(Perhaps more simply, the conditional distribution
of $\cal S$ given ${\cal A}=\alpha$ and ${\cal G}=\Gamma$ is $\Gamma_\alpha(s)$, which depends on $\alpha$ but is 
independent of $\Gamma$.) The general result for a Markov chain then gives for the present case
\be
I({\cal S};{\cal A})=I({\cal S};{\cal A}|{\cal G})+I({\cal S};{\cal G}).
\label{comprel}
\ee
This equation is among the main results of this paper. It says that the complexity of the MAS results from 
(the metastate average of) that of a Gibbs state, plus that of the metastate itself: $K_\rho=K_\Gamma+K_\kappa$,
an intuitively appealing statement (we emphasize that a $\kappa$ expectation is 
included in the definition of each one). It should not be confused with the simple chain rule; the 
Markov property was crucial to reduce $I({\cal S};{\cal A},{\cal G})$ to $I({\cal S};{\cal A})$. 
It implies inequalities as before; of the three complexities, the largest is that of the MAS, $K_\rho$. 
Either or both of the terms on the right could be zero, and both will be at high temperature. 

\subsubsection{Monotonicity and bound by mutual information with nearby spins}
\label{subsubmono}

Now and for the remainder of the paper, we usually consider the mutual information for ${\cal S}_\Lambda$
rather than for $\cal S$, and sometimes take the $\Lambda\to\infty$ limit; $\Lambda$ could be an arbitrary finite set, but 
one may think of the example of a hypercube 
$\Lambda_W$ of side $W$, centered at the origin (we take $W$ odd), with sides parallel to the coordinate axes of ${\bf Z}^d$.

For this case, we may first notice a general result in the setting of Gibbs states and metastates. Suppose that 
$\Lambda_1\subseteq\Lambda_2$. Then we can identify 
${\cal S}_{\Lambda_2}$ with the pair $({\cal S}_{\Lambda_1},{\cal S}_{\Lambda_2-\Lambda_1})$, and the general inequalities
that result from the chain rule imply, for example, that
\be
I({\cal S}_{\Lambda_1};{\cal A}|{\cal G})\leq I({\cal S}_{\Lambda_2};{\cal A}|{\cal G})
\ee
and similarly for the other complexities relativized to the windows $\Lambda_1\subseteq\Lambda_2$. Hence the mutual information 
is a  monotonically increasing function of the window $\Lambda$.

We further notice that the random variables $\cal A$ and $\cal G$ reflect dependence on spins or bonds far away, effectively 
at infinity. We expect that the mutual information of ${\cal S}_\Lambda$ with either of them must be mediated by the spins 
in-between, due to the finite range of the interactions. That is, we should have a refined Markov chain, in the form
\be
{\cal G}\to{\cal A}\to{\cal S}_{\Lambda^c}\to{\cal S}_\Lambda,
\ee
or we could omit $\cal A$ or $\cal G$. (The displayed form means that any three successive terms form a Markov chain 
as above.) Indeed, the Markov property does hold at ${\cal S}_{\Lambda^c}$, because
for any Gibbs state (for the given $J$) the conditional distribution of ${\cal S}_\Lambda$ given 
${\cal S}_{\Lambda^c}$ is given by the fixed formula eq.\ (\ref{gibbsdef}) independent of which
Gibbs (or pure) state is considered. [Eq.\ (\ref{gibbsdef}) is discussed further for the long-range case
in Appendix \ref{app:gibbs}.]

Consequently, for each complexity that we consider, we have an inequality, for example
\be
K_\rho(\Lambda)\leq I({\cal S}_\Lambda;{\cal S}_{\Lambda^c}),
\label{Ilamblamc}
\ee
that results directly from the pipeline inequality above where, to be completely explicit, the right-hand side is the 
mutual information in the MAS, $ I({\cal S}_\Lambda;{\cal S}_{\Lambda^c})=I({\cal S}_\Lambda;{\cal S}_{\Lambda^c})_\rho$.
The same upper bound holds for the other complexities $K_\Gamma(\Lambda)$, $K_\kappa(\Lambda)$ also. 
Hence it will be sufficient to upper-bound the right-hand side of this expression, and we will do so for its disorder average over 
$J$. 

In a strictly short-range model, we can replace $\Lambda^c$ in these formulas by $\Lambda_2-\Lambda$ for a sufficiently
large finite set $\Lambda_2$, with $\Lambda\subseteq\Lambda_2$, such that the Markov property still holds. For Ising or other 
discrete spins, this gives one way to recover the bound on any of the complexities by a (model-dependent) 
constant times the surface area of $\Lambda$, thus generalizing the result of HR to the complexity of the metastate in addition 
to the other two.

\subsubsection{Bounds on expected mutual information of a partition}
\label{subsubbounds}

In light of the inequality (\ref{Ilamblamc}), we will now consider the
mutual information between the spins in two parts of our system. That is, we partition the sites
into two or more parts, and we will find upper bounds on the expectation over $J$ of the mutual
information between the spins in the two chosen parts. This can be done straightforwardly for a finite
system, and then we can consider infinite size. We first consider three useful finite-size examples.

We consider finite systems with sites in a finite set, say $\Lambda''$. For an arbitrary Hamiltonian $H''=H''(s|_{\Lambda''})$, 
say, we will define the free energy for a sum over a subset $\Lambda'^c\subseteq\Lambda''$ of the spin variables on which $H''$ depends 
($\Lambda'=\Lambda''-\Lambda'^c$),
\be
e^{-F_{H''}(s|_{\Lambda'})/T}=\sum_{s|_{\Lambda'^c}}e^{-H''(s|_{\Lambda''})/T},
\ee
so the free energy $F_{H''}(s|_{\Lambda'})$ will usually depend on $s|_{\Lambda'}$. If $\Lambda'=\emptyset$, 
we write $F_{H''}$ for $F_{H''}(\emptyset)$; often $\Lambda''=\Lambda$, the full system, and then we write $F$ for $F_H$,
where $H$ is the full Hamiltonian on $\Lambda$. 

First, consider the simplest case of a system with two parts, in which the two parts are $\Lambda_1$, $\Lambda_2$, with 
$\Lambda_1\cup\Lambda_2=\Lambda$, $\Lambda_1\cap\Lambda_2=\emptyset$. We will write $s=s|_\Lambda$ as before, 
but write $s|_{\Lambda_1}$, $s|_{\Lambda_2}$ as $s_1$, $s_2$ for brevity (no confusion with the strict notation that $s_i$ 
is the spin at a single site should arise). 
We simplify the notation for finite-size Gibbs distributions, as follows. For the two-part system,
we have
\be
p(s_1,s_2)=e^{-H/T+F/T}
\ee
where $H=H(s)$ is the Hamiltonian of the whole system, and the marginal distribution $p(s_1)=\sum_{s_2}p(s_1,s_2)$ is
\be
p(s_1)=e^{-H_1/T-F_{H-H_1}(s_1)/T+F/T},
\ee
and similarly for $p(s_2)$ by replacing $1$ with $2$. Here $H_1=H_1(s_1)$ is the Hamiltonian $H$ with all terms
involving spins in $\Lambda_2$ omitted, or alternatively it can be thought of as $H$ projected into $\Lambda_1$
by summing $H$ over $s_2$, as that annihilates the unwanted terms as in the Fourier-Walsh expansion (recall that the constant 
term in the Hamiltonian is always set to zero). Then the mutual information of the two parts is
\be
I(1;2)=\sum_{s_1,s_2}p(s_1,s_2)\ln\frac{p(s_1,s_2)}{p(s_1)p(s_2)}.
\ee

Now consider the same system with all terms that involve spins in both $\Lambda_1$ and $\Lambda_2$ set to zero, so the parts
are decoupled. Denote the corresponding Gibbs distributions by $p^{(0)}(s_1,s_2)$, $p^{(0)}(s_1)$, and $p^{(0)}(s_2)$,
so $p^{(0)}(s_1,s_2)=p^{(0)}(s_1)p^{(0)}(s_2)$. Multiplying and dividing by the latter function inside the logarithm gives
\bea
I(1;2)&=&D[p(s_1,s_2)||p^{(0)}(s_1,s_2)]-D[p(s_1)||p^{(0)}(s_1)]\non\\
&&{}-D[p(s_2)||p^{(0)}(s_2)]\\
&\leq& D[p(s_1,s_2)||p^{(0)}(s_1,s_2)]
\eea
by non-negativity of relative entropy $D$. The right-hand side is 
\bea
\lefteqn{D[p(s_1,s_2)||p^{(0)}(s_1,s_2)]=}\qquad&&\non\\
&=&-\frac{1}{T}\sum_{s_1,s_2}p(s_1,s_2)(H-H_1-H_2)\non\\
&&{}+\frac{F-F_{H_1}-F_{H_2}}{T}
\eea
Here $F_{H_1}$ is defined as above, and is independent of $s_2$, and similarly for $F_{H_2}$; this allowed 
the sum over $s_1$, $s_2$ in the second term to be carried out already.

Next we take the expectation ${\bf E}$ of the preceding $D$ over $J$ using the distribution $\nu(J)$ which was specified in Sec.\ 
\ref{sgmodels}. In fact, we only need 
to take the expectation ${\bf E}'$ with respect to the terms that involve spins in both parts $1$ and $2$, in other words $X$ such that 
$X\cap\Lambda_1\neq\emptyset$, $X\cap\Lambda_2\neq\emptyset$, as that will produce the desired bounds which 
do not depend on the other bonds. The first term in $D$ contains $H-H_1-H_2$, which consists precisely of all the terms that 
involve both parts. Its expectation is
\be
\frac{1}{T}{\bf E}'\sum_{X\subseteq\Lambda:X\cap\Lambda_1\neq\emptyset,X\cap\Lambda_2\neq\emptyset}J_X\langle s_X\rangle
\ee
[the expectation is in the Gibbs state $p(s_1,s_2)$]. 

Here it will be useful to introduce a version of a technique (see e.g.\ Ref.\ 
\cite{cg_book}) that will be used repeatedly. For given $X$, consider the modified Hamiltonian with $-J_Xs_X$ replaced by 
$-\lambda J_X s_X$ to form the interpolating Hamiltonian $H-(\lambda-1)J_X s_X$, and examine what happens as $\lambda$ 
changes from $0$ to $1$. We will write $\langle\cdots\rangle_\lambda$ for the thermal average in the interpolating finite-size Gibbs
distribution ($\langle\cdots\rangle=\langle\cdots\rangle_{\lambda=1}$ will continue to mean the usual full thermal average with $H$). 
Then we have
\be
{\bf E}'J_X\langle s_X\rangle={\bf E}'J_X\int_0^1d\lambda\, \frac{d}{d\lambda}\langle s_X\rangle_\lambda,
\ee
where the term from the lower limit $\lambda=0$ is zero because the $J_X$s are independent, which implies that
the thermal average $\langle s_X\rangle_{\lambda=0}$ is independent of $J_X$, and centered. Then for Ising spins
\bea
{\bf E}'J_X\langle s_X\rangle&=&\frac{1}{T}{\bf E}'J_X^2\int_0^1d\lambda\,\left(1-\langle s_X\rangle_\lambda^2\right)\\
&\leq&\frac{1}{T}{\bf E}'J_X^2\\
&=&\frac{1}{T}{\rm Var}\,J_X.
\label{nonGibp}
\eea
(For Gaussian bonds, this can also be obtained by integration by parts, without introducing $\lambda$ at this stage.
The same bound for $m$-vector spins with $(X,x)$ in place of $X$ is obtained similarly. An alternative to this upper bound
is to use ${\bf E}'J_X\langle s_X\rangle\leq {\bf E}'|J_X|$, which still leads to useful results in the short-range case; this variation can
also be made in the following calculations.)

Hence the first term in ${\bf E}D$ is bounded above by
\be
\frac{1}{T^2}\sum_{X\subseteq\Lambda:X\cap\Lambda_1\neq\emptyset,X\cap\Lambda_2\neq\emptyset}{\rm Var}\,J_X.
\ee
For $m$-vector spins we obtain similarly an upper bound on the first term in $D$, which is the same except that the sum over 
$X$ becomes a sum over $(X,x)$. 

Under our current condition that all $J_X$ are centered, the second term in ${\bf E}D$ is in fact non-positive and can be dropped 
to obtain the upper bound on ${\bf E}D$ 
and on ${\bf E}I(1;2)$, as follows. It is the difference of the free energies $F$, with bonds between $1$ and $2$ included,
and $F_{H_1}+F_{H_2}$ with those bonds omitted (set to zero). For the interpolating Hamiltonian $H+(\lambda-1)(H-H_1-H_2)$, 
the corresponding free energy $F(\lambda)=F_{H+(\lambda-1)(H-H_1-H_2)}$ obeys $F(1)=F$, $F(0)
=F_{H_1}+F_{H_2}$, and is a concave function of $\lambda$ (its second derivative with respect to $\lambda$ is $\leq0$). Note that 
$F(1)-F(0)=\int_0^1d\lambda\, dF(\lambda)/d\lambda$. The first derivative of $F(\lambda)$, taken at $\lambda=0$, 
is the thermal average of $H-H_1-H_2$ in the Gibbs distribution $p^{(0)}(s_1,s_2)$ in which the couplings between $1$ and $2$ 
are set to zero, so again the ${\bf E}'$ expectation of it is zero, as the $J_X$ have mean zero and are independent; then by concavity 
of $F(\lambda)$ and hence of ${\bf E}F(\lambda)$, the first derivative of ${\bf E}F(\lambda)$ is non-positive for all $\lambda\geq0$. 
It follows that ${\bf E}F(\lambda)\leq {\bf E}F(0)$ for all $\lambda\geq0$, so ${\bf E}(F-F_{H_1}-F_{H_2})\leq0$, which is what 
we set out to prove. [If the $J_X$s are not centered, then the preceding upper bound $\leq 0$ can be replaced by
$\leq|{\bf E}J_X|\leq{\bf E}|J_X|$, summed over the $X$ that involve spins in both $1$ and $2$.]

Thus we have proved that
\be
{\bf E} I(1;2)\leq \frac{1}{T^2}\sum_{X\subseteq\Lambda:X\cap\Lambda_1\neq\emptyset,X\cap\Lambda_2
\neq\emptyset}{\rm Var}\,J_X,
\label{bound1}
\ee
for the Ising case, with the usual change $X\to(X,x)$ for more general models. This is another of the main results of this paper, and 
similar formulas will be found more generally, not just for the case considered here. Note that, from the derivation, the upper bound
was a bound on (minus) the expectation of the ``surface energy'' (the thermal average of $H-H_1-H_2$), divided by $T$; 
compare Ref.\ \cite{vEvH_85}. We can already take $\Lambda_1=\Lambda_W$ for $W$ fixed, and let
$\Lambda$ increase to infinity, and then we expect the corresponding limit of the upper bound on ${\bf E}I({\cal S}_{\Lambda_W};
{\cal S}_{\Lambda_W^c})$ to hold for the limit; we will return to this later. 

Next we consider the more general case of a partition into three disjoint regions $\Lambda_1$, $\Lambda_2$, $\Lambda_3$, 
and bound the mutual information between two of them, say regions $1$ and $2$. We choose to compare with the Gibbs 
distributions in which all interactions involving spins in both $\Lambda_1$ and $\Lambda_2\cup\Lambda_3$ 
are set to zero; then we have again
\be
I(1;2)\leq D[p(s_1,s_2)||p^{(0)}(s_1,s_2)],
\ee
where now, with $p^{(0)}(s_1,s_2)=p^{(0)}(s_1)p^{(0)}(s_2)$,
\bea
p(s_1,s_2)&=&e^{-H_{12}/T-F_{H-H_{12}}(s_1,s_2)/T+F/T},\\
p^{(0)}(s_1)&=&e^{-H_1/T+F_{H_1}/T},\\
p^{(0)}(s_2)&=&e^{-H_2/T-F_{H_{23}-H_2}(s_2)/T+F_{H_{23}}/T}.
\eea
Here $H_{ij}$, for $i$, $j\in\{1,2,3\}$, are $H$ projected so that it depends only on $s_i$, $s_j$, $H_i$ are $H$ projected 
so it depends on $s_i$ only, and the $F$s depend only on the spin variables displayed. This gives
\bea
I(1;2)&\leq&\frac{1}{T}\sum_{s_1,s_2}p(s_1,s_2)\left[-(H_{12}-H_1-H_2)\right.\non\\
&& \left.{} -(F_{H-H_{12}}(s_1,s_2)-F_{H_{23}-H_2}(s_2)) \right]         \non\\
&&{}+\frac{F-F_{H_1}-F_{H_{23}}}{T}
\eea
Taking the expectation value, the first term is bounded above by almost the same sum as when $\Lambda_3$ was empty, except 
that now $X\subseteq\Lambda$ is replaced by $X\subseteq\Lambda_1\cup\Lambda_2$, and the third term is again non-positive 
and can be dropped. 

The second term contains 
\be
F_{H-H_{12}}(s_1,s_2)-F_{H_{23}-H_2}(s_2),
\ee
which again is a free energy, this time with $s_1$, 
$s_2$ fixed, minus the same quantity with the bonds that connect $1$ to $3$ removed, so $3$, where the summation over $s_3$ occurs, 
is completely decoupled from $1$. As this combination occurs with a negative overall sign, we cannot use the same argument to drop its
expectation, and must find a bound on its expectation from the other side instead. We can write this free energy difference as the integral 
of the derivative of $F(\lambda)=
F_{H-H_{12} + (\lambda-1)(H-H_{12}-H_{23}+H_2)}$, where dependence on $s_1$, $s_2$ is implicit. The $\lambda$-dependent 
term of the interpolating Hamiltonian $H-H_{12} + (\lambda-1)(H-H_{12}-H_{23}+H_2)$
contains only the interactions that involve spins in both $\Lambda_1$ and $\Lambda_3$, and possibly also some in $\Lambda_2$.
In writing those terms, we will use notation $s_X=s_{X^{(1)}}s_{X^{(2)}}s_{X^{(3)}}$ (for Ising spins) which shows how $s_X$, 
and $X$ itself, is split between the three parts ($X^{(i)}=X\cap\Lambda_i\subseteq\Lambda_i$); $X^{(2)}$ could be empty, and 
$s_\emptyset=1$, of course. 

The minor subtlety with finding the upper bound is that the outer thermal average depends on the bonds whose strength we vary
in $F(\lambda)$, though they do not vary in the outer average. We have
\bea
\lefteqn{-\frac{1}{T}\frac{dF}{d\lambda}=}\qquad&&\\
&&\frac{1}{T}\sum_{X\subseteq\Lambda: X^{(1)}\neq\emptyset,X^{(3)}\neq\emptyset}
J_X s_{X^{(1)}} s_{X^{(2)}}\left \langle s_{X^{(3)}}\right\rangle_{s_1,s_2,\lambda},\non
\eea
where the average $\langle\cdots \rangle_{s_1,s_2,\lambda}$ is taken using the conditional Gibbs distribution for the interpolating 
Hamiltonian, with parameter $\lambda$, and $s_1$ and $s_2$ fixed, not summed over,
as indicated. Then in order to bound the ${\bf E}'$ expectation, in the $X$th term in the $X$ sum we must introduce a second 
parameter $\lambda'$
into $-J_X s_X$ only, with an integral from $0$ to $1$ of the $\lambda'$ derivative, in both the outer thermal average and the inner one; 
in the latter, this means replacing $\lambda$ with $\lambda\lambda'$ in the $X$th term of the interpolating Hamiltonian. This leads to 
\bea
\lefteqn{-\frac{1}{T}{\bf E}\left\langle\frac{dF}{d\lambda}\right\rangle}&&\non\\
&=&\frac{1}{T^2}{\bf E}{\sum_X}'J_X^2\int_0^1d\lambda'\,\left[\lambda\left(1-\left\langle\left\langle s_{X^{(3)}}
\right\rangle_{s_1,s_2,\lambda,\lambda'}^2\right\rangle_{\lambda'}\right)\right.\non\\
&&\qquad{}+\left\langle s_{X^{(3)}}\left\langle s_{X^{(3)}}\right\rangle_{s_1,s_2,\lambda,\lambda'}\right\rangle_{\lambda'}\non\\
&&\left.\qquad{}-\left\langle s_X \right\rangle_{\lambda'}\left\langle s_{X^{(1)}} s_{X^{(2)}}\left\langle s_{X^{(3)}}
\right\rangle_{s_1,s_2,\lambda,\lambda'}\right\rangle_{\lambda'} \vphantom{\left(1-\left\langle s_{X^{(3)}}
\right\rangle_{s_1,s_2,\lambda,\lambda'}^2\right)}\right]\\
&\leq&\frac{1}{T^2}\sum_{X\subseteq\Lambda: 
X^{(1)}\neq\emptyset,X^{(3)}\neq\emptyset}{\rm
 Var}\,J_X[\lambda+2],
\eea
where in the first line the sum $\sum_X'$ is over the same $X$ as in the last line.
Then integrating this bound from zero to one, we obtain
\be
-\frac{1}{T}{\bf E}\langle[F(1)-F(0)]\rangle\leq \frac{5}{2T^2}\sum_{X\subseteq\Lambda: 
X^{(1)}\neq\emptyset,X^{(3)}\neq\emptyset}{\rm Var}\,J_X,
\ee
so finally we have the desired bound
\bea
{\bf E} I(1;2)&\leq& \frac{1}{T^2}\sum_{X\subseteq\Lambda_1\cup\Lambda_2:X\cap\Lambda_1\neq\emptyset,X\cap\Lambda_2
\neq\emptyset}{\rm Var}\,J_X\quad\non\\
&&{}+\frac{5}{2T^2}\sum_{X\subseteq\Lambda: X\cap\Lambda_1\neq\emptyset,X\cap\Lambda_3
\neq\emptyset}{\rm Var}\,J_X.
\label{bound2}
\eea
Again, the same bound (with changes as before) is obtained for $m$-vector spins. 
Essentially, the first term is a sum of interactions between $1$ and $2$ only, while the second is a sum of interactions
that involve $1$ and $3$ and possibly also $2$ (the latter only for $p\geq3$ interactions).

The preceding bound can be applied in more than one way. If $1$ and $2$ are adjacent regions, while $3$ is distant from $1$, 
then as the distance from $1$ to $3$ goes to infinity, those interactions should go to zero to satisfy the convergence condition,
and it reduces to the same bound as when there were only two parts. If instead $3$ is close to $1$, and $2$ goes off to infinity, then 
the present bound is weaker, so the other case gives a better bound. 

As a third example of a bound on mutual information, we consider the conditional mutual information between two parts $1$ and $2$, 
conditioned on $3$, $I(1;2|3)$. The application we have in mind is to regions $1$ and $2$ far apart, and separated by part $3$, and
we expect that $1$ and $2$ become conditionally independent in the limit. Thus we consider 
\be
I(1;2|3)=\sum_{s_1,s_2,s_3}p(s_1,s_2,s_3)\ln\frac{p(s_1,s_2|s_3)}{p(s_1|s_3)p(s_2|s_3)}.
\ee
In this case we can compare with the Gibbs distribution with the interaction terms that involve at least one spin in each of $1$ and $2$
set to zero, because the corresponding conditional probability factorizes, $p^{(0)}(s_1,s_2|s_3)=p^{(0)}(s_1|s_3)p^{(0)}(s_2|s_3)$.
As before, we then have
\be
I(1;2|3)\leq D[p(s_1,s_2|s_3)||p^{(0)}(s_1,s_2|s_3)],
\ee
where now
\bea
p(s_1,s_2|s_3)&=&e^{-(H-H_3)/T+F_{H-H_3}(s_3)/T},\\
p^{(0)}(s_1|s_3)&=&e^{-(H_{13}-H_3)/T+F_{H_{13}-H_3}(s_3)/T},\\
p^{(0)}(s_2|s_3)&=&e^{-(H_{23}-H_3)/T+F_{H_{23}-H_3}(s_3)/T}.
\eea
The calculation of the upper bound on the expectation of this $D$ is similar to the previous example, and the result is
\be
{\bf E}I(1;2|3)\leq \frac{3}{T^2}\sum_{X\subseteq\Lambda: X\cap\Lambda_1\neq\emptyset,X\cap\Lambda_2
\neq\emptyset}{\rm Var}\,J_X.
\label{bound3}
\ee
(In this case, a term that resembles the one that produced the $1/2$ before now has the opposite sign, so gives zero in the upper 
bound instead; the other terms combine to give the factor $3$.) Thus if the direct interactions involving both $1$ and $2$ 
are weak, the conditional mutual information is small, because for fixed spins in $3$, the mutual information or correlation arises only
from the direct interactions. 

So far in this subsection, we have produced upper bounds on mutual information in finite-size systems. We still have to obtain
bounds for infinite size, to make contact with the previous discussion of complexity. If we were concerned only with 
{\em short-}range systems, there would not be much of a problem. In that case, if a Gibbs state (say, the MAS $\rho$) 
that depends on the bonds $J$ is given in infinite size, then the preceding bounds continue to hold, 
because the short-range condition ensures absolute convergence of the relevant sums in infinite size, and in particular allows 
the term-by-term use of bound (\ref{nonGibp}). (We discuss this further, though with different motivation, and showing how to
avoid reference to the free energy of an infinite portion of the system during the derivation, in Appendix 
\ref{app:trivi}. Thus here we are {\em not} saying ``take the thermodynamic limit'', as the existence
of a limit of the finite-size Gibbs states is not clear even in the strictly short-range case, and that is the reason for introducing 
metastates.) Then the same upper bound (\ref{bound1}) with $\Lambda_1=\Lambda$, $\Lambda_2=\Lambda^c$ [or the limit 
$\Lambda_2\to\Lambda^c$ of bound (\ref{bound2})] applies to the expected mutual information of ${\cal S}_\Lambda$ and 
${\cal S}_{\Lambda^c}$ in $\rho$, that is to ${\bf E}I({\cal S}_\Lambda;{\cal S}_{\Lambda^c})_\rho=
{\bf E}I({\cal S}_\Lambda;{\cal S}_{\Lambda^c})$. By the Markov chain arguments, the same upper bound also applies 
to the expected complexity of the MAS, giving
\be
{\bf E}K_\rho(\Lambda)\leq \frac{1}{T^2}\sum_{X\in{\cal X}:X\cap\Lambda\neq\emptyset,X\cap\Lambda^c
\neq\emptyset}{\rm Var}\,J_X.
\label{boundfinal}
\ee 
As the complexity of the MAS is at least as large as either of the other two complexities $K_\Gamma(\Lambda)$, $K_\kappa(\Lambda)$ 
by relation (\ref{comprel}), the same upper bound as in (\ref{boundfinal}) applies
to the $\nu$-expectation of those also.

But in fact we wish to include the long-range cases, and then we need to take some limits carefully (some readers 
may prefer to skip the following more technical discussion). Again, it is useful to consider the expected 
complexity of the MAS. Even in finite size, if we average the state over disorder $J_X$ such that $X$ has zero intersection with 
$\Lambda_1\cup\Lambda_2$ [in the notation of the bound (\ref{bound2}) above], additional effective interactions with the 
spins in region $1$ are generated (the effective Hamiltonian can be obtained by applying Fourier-Walsh expansion to
the logarithm of the average state), so a bound may not take the same form as before. It would be better to remove the interactions
of region $1$ with the outer region before proceeding with the average and the infinite-size limit. The most efficient way to do so 
is to use the truncated-interaction metastate $\kappa_{\rm ext}$ introduced in the course of the proof of Proposition 3 in 
Appendix \ref{app:gibbs}. First, in finite size, we 
consider Gibbs states $\Gamma^{n'}_{(\Lambda,\Lambda')}$, where $n'$ refers 
to a finite system on $\Lambda_{n'}'$ ($=\Lambda_L$, say, for now) and $\Lambda$, $\Lambda'$ are smaller finite sets of spins, 
with $\Lambda\subseteq\Lambda'\subseteq\Lambda'_{n'}$ (thus we will now revert to the notation of the preceding subsections). 
This is the Gibbs state for the usual Hamiltonian on $\Lambda_{n'}'$ 
but with all terms that involve spins in both $\Lambda$ and $\Lambda'^c$ dropped. For $\Lambda_R$ such that 
$\Lambda'\subseteq\Lambda_R\subseteq\Lambda_{n'}'$, we now take the average of $\Gamma^{n'}_{(\Lambda,\Lambda')}$ over 
disorder (partially) outside $\Lambda_R$ using $\nu(J|_{{\cal X}(\Lambda_{n'}')-{\cal X}(\Lambda_R)})$ to obtain 
$\rho_{(\Lambda,\Lambda')R}^{n'}$. The average state $\rho_{(\Lambda,\Lambda')R}^{n'}$
can be expressed as a Gibbs state for an effective Hamiltonian, and the latter can be obtained by using the Fourier-Walsh expansion 
of $\ln \rho_{(\Lambda,\Lambda')R}^{n'}$. Due to the truncated interaction, the terms in the effective Hamiltonian that involve 
the spins in $\Lambda$ are unaffected by the averaging; they are the same as in $H$. Then we can bound the expected 
mutual information between $\Lambda$ and $\Lambda'-\Lambda$ in the state $\rho_{(\Lambda,\Lambda')R}^{n'}$ exactly 
as in the bound (\ref{bound2}) above, where $\Lambda$ and $\Lambda'-\Lambda$ replace $\Lambda_1$ and $\Lambda_2$, and the term 
involving interactions with region $3$ (i.e.\ $\Lambda_3=\Lambda_{n'}'-\Lambda'$) drops out as those interactions are zero. 

Next, as $\Lambda_{n'}'\to\infty$, that is as $n'\to\infty$, and then $R\to\infty$ (along the subsequences that produce the extended 
or truncated-interaction metastate), $\rho_{(\Lambda,\Lambda')R}^{n'}$ tends to the MAS $\rho_{(\Lambda,\Lambda')}$ 
\cite{ns96b,ns_rev,aw} in distribution \cite{chung_book,breiman_book};
$\rho_{(\Lambda,\Lambda')}$ is the average under the extended metastate of an infinite-size state $\Gamma_{(\Lambda,\Lambda')}$ 
with truncated interactions, drawn from the extended metastate. The upper bound is still finite when $n'\to\infty$, so as in the 
proof of Proposition 2 in Appendix \ref{app:gibbs} we have an upper bound on the expected 
mutual information ${\bf E}I({\cal S}_\Lambda;{\cal S}_{\Lambda'-\Lambda})_{\rho_{(\Lambda,\Lambda')}}$ in infinite size 
[where in this paragraph ${\bf E}$ means expectation under $\nu(J)$]. In the proof of Proposition 3, it was proved that in any 
given sample of the indexed set of $(\Gamma,(\Gamma_{(\Lambda,\Lambda')})_{(\Lambda,\Lambda')})$ [that is, a state 
$\Gamma$ without truncation, and the collection of states with interactions truncated for all pairs $(\Lambda,\Lambda')$] drawn 
(simultaneously) from the extended metastate, $\Gamma_{(\Lambda,\Lambda')}\to\Gamma$ in the strongest possible sense 
as $\Lambda'\to\infty$ (along a certain sequence $\Lambda_n$) for any $\Lambda$, $\nu\kappa_{\rm ext}$-almost surely.
It then follows that also $\rho_{(\Lambda,\Lambda')}\to\rho$. Then again the same upper bound applies to the expected 
mutual information of ${\cal S}_\Lambda$ and ${\cal S}_{\Lambda^c}$ in $\rho$, and hence the same upper bound as in
(\ref{boundfinal}) applies to the three complexities ${\bf E}K_\Gamma(\Lambda)$, ${\bf E}K_\kappa(\Lambda)$, 
${\bf E}K_\rho(\Lambda)$ in all the finite-range models. These are the final forms of the three inequalities, and are 
among the main results of this paper.

\subsubsection{Other distributions for disorder}
\label{subsubothdis}

So far in this paper, we have considered only random variables $J_X$ that are independent with mean zero
(except that we can allow nonzero mean when $|X|=1$) and which all have the same distribution when rescaled by 
$\sqrt{{\rm Var}\,J_X}$; the discussion also applies to $J_{(X,x)}$ for more general
models with the usual modifications. Here we will briefly mention the extension to some other
distributions. (Elsewhere in the paper, we continue to use the models of Sec.\ \ref{sgmodels}.) An easy extension of 
the results is to the case of $J_X$ with a distribution that is a weighted mixture (i.e.\ a convex combination) of, say, Gaussian 
distributions, independently for each $X$. For these, it can easily be seen that the complexity upper bounds expressed in terms of 
${\rm Var}\,J_X$ are unchanged, because each such variance is just the weighted sum of variances of the Gaussians in the mixture. 

In particular, the use of diluted bonds is popular in simulations \cite{lprtrl}. 
In these models, each $J_X$ is either a Gaussian random variable of mean zero and variance $1$, with weight (probability) $p_X$, 
or zero with probability $1-p_X$, and the $J_X$s for each $X$ are independent.  
A $\delta$-function distribution at zero can be viewed as a Gaussian with zero variance. Then for the $d=1$ $p=2$ 
power-law model, one can take $p_{ij}=|i-j|^{-2\sigma}$, so that ${\rm Var}\,J_{ij}$ takes its usual form $\sim|i-j|^{-2\sigma}$. 
These models are expected to be in the same universality class for the behavior at $T=T_c$ as the previous power-law model, 
and are expected to exhibit similar behavior also more generally, for the range $1/2<\sigma<1$ (though not when $T=0$ 
and $\sigma>1$ \cite{read18}). Then as we said, the upper bounds on the expected complexities take the same form as in the 
usual model also. 

We note, however, that in these diluted-bond models the sums like $\sum_X |J_Xs_X|=\sum_X |J_X|$ converge if 
$\sum_X p_X$ does, and the latter is the same as the corresponding sum of variances. Hence the sums of interest (for e.g.\ 
the existence of Gibbs states) converge {\em absolutely} whenever the convergence condition (\ref{cvgce_cond}) holds, and 
not only when the more restrictive condition (\ref{sh_cvgce_cond}) holds. Consequently, for the purposes of this paper the 
diluted models can be handled with the easier methods that apply to the short-range models, even in cases like the $d=1$ 
$p=2$ models with $1/2<\sigma<1$ that for the models of Sec.\ \ref{sgmodels} we earlier classed as long-range. This 
illustrates again that, while the condition (\ref{cvgce_cond}) is the important one in general, the technical distinction between 
short- and long-range cases, which was based on absolute convergence properties of sums, may not be so important physically.

More generally, if the scaling assumption on the distributions is dropped, bounds based on sums of ${\bf E}|J_X|/T$ in 
place of ${\rm Var}\,J_X/T^2$ can be used, as mentioned in Sec.\ \ref{subsubbounds} (for the diluted models, this gives 
the same result just mentioned), and may be more effective, for example for very broad distributions where ${\rm Var}\,J_X$ 
may be infinite or its sum over $X$ may be poorly convergent. This leads to the following general form of disorder for which
our results on complexity and existence of Gibbs states hold: the bond $J_X$ for each $X$ can be a sum $J_X=J_X^{(1)}+ J_X^{(2)}$, 
where $\left(J_X^{(1)}\right)_X$, $\left(J_X^{(2)}\right)_X$ are all independent, the joint distribution is homogeneous (i.e.\ 
translation invariant), the $\left(J_X^{(1)}\right)_X$ are not necessarily centered but obey the condition for absolute convergence,
\be
\lim_{|\Lambda'|\to\infty} \sum_{p\geq1} \sum_{X\subseteq \Lambda':i\in X,|X|=p}
{\bf E}|J_{X}^{(1)}|<\infty,
\ee 
while $\left(J_X^{(2)}\right)_X$ are centered and obey the convergence condition
\be
\lim_{|\Lambda'|\to\infty} \sum_{p\geq1} \sum_{X\subseteq \Lambda':i\in X,|X|=p}
{\rm Var}\,J_{X}^{(2)}<\infty.
\ee 
(These forms of disorder resemble, but the conditions are somewhat more restrictive than, those used in Ref.\ \cite{zegarlinski_91} 
for thermodynamics, and are mentioned there; see also Appendix \ref{app:energy} in this paper.) Then in the upper bounds in 
the proofs of results, the corresponding form of bound is applied to terms containing $J_X^{(1)}$ or $J_X^{(2)}$, respectively, 
which results in simple modifications to the bounds on complexity; they become
\bea
{\bf E}K_\rho(\Lambda)&\leq&\frac{2}{T}\sum_{X\in{\cal X}:X\cap\Lambda\neq\emptyset,X\cap\Lambda^c
\neq\emptyset}{\bf E}\,|J_X^{(1)}|\non\\
&&{}+ \frac{1}{T^2}\sum_{X\in{\cal X}:X\cap\Lambda\neq\emptyset,X\cap\Lambda^c
\neq\emptyset}{\rm Var}\,J_X^{(2)},\qquad
\label{boundfinal'}
\eea
and the same bound for the other complexities. We note that this form applies even when bonds with $p>1$ have nonzero mean,
so more generally than other results of this paper, and includes models without disorder as special cases.
For $m$-vector models, we can also require that the Hamiltonian restricted to any finite set $\Lambda$ of spins be almost-surely 
finite for all $s|_\Lambda$, as before. 

There are also versions of Theorem 2 (see Appendix \ref{app:trivi}) for one-dimensional models with this more general form of disorder.
In place of a short-range model, we take bonds with $J_X^{(2)}=0$ for all $X$. Then we write $J_X^{(1)}={\bf E}J_X^{(1)}
+(J_X^{(1)}-{\bf E}J_X^{(1)})$, and now a similar argument as in the proof of Theorem 2 shows that a sufficient 
condition for the Gibbs state at $T>0$ to be unique is that both
\be
\sum_{X\in{\cal X}:X\cap{\bf Z}_-\neq\emptyset,X\cap{\bf Z}_+\neq\emptyset}|{\bf E}\,J_X^{(1)}| < \infty
\label{domwall1}
\ee
and 
\be
\sum_{X\in{\cal X}:X\cap{\bf Z}_-\neq\emptyset,X\cap{\bf Z}_+\neq\emptyset}{\rm Var}\,J_X^{(1)} < \infty
\label{domwall2}
\ee
hold. The second of these is the same as the condition in the original version of Theorem 2, while the first has the same form
as the condition in a non-random model \cite{ruelle}, which is thus a special case of this result. 
An alternative sufficient condition
is to replace that above by the single condition that the same sum, but now of ${\bf E}|J_X^{(1)}|$, be finite; this may be weaker 
if the distribution of each $J_X^{(1)}$ is very broad. There are also other valid conditions that combine both forms.

\subsubsection{Discussion}
\label{subsubdisc}

As basic examples, if we use the upper bound (\ref{boundfinal}) for $\Lambda=\Lambda_W$ 
a hypercube of side $W$, for short-range models 
the sum gives a result proportional to the surface area $\propto W^{d-1}/T^2$ (for fixed $T>0$ and for a strictly short-range 
model, this is similar to Ref.\ \cite{hr}). For long-range models, it can be viewed as a definition of surface area for these cases, 
and generally grows faster than $W^{d-1}/T^2$ as $W$ increases. For the example of the one-dimensional $2$-spin power law 
model, which is long-range when $1/2<\sigma\leq 1$, the sum behaves as $W^{2-2\sigma}/T^2$ (for $\sigma<1$), as stated 
in the introduction. (The same sum playing the role of the surface area of the window also arose in another distinct bound in 
the Appendix of Ref.\ \cite{read18}.) In these models, for $\sigma>1$, the bound on any of the complexities is order one as 
$W\to\infty$, indicative of low complexity, and consistent with the absence of a spin glass phase at $T>0$ in these models. 
We consider these cases further in Appendix \ref{app:trivi}.

We also comment that the third bound (\ref{bound3}), where we take $\Lambda_3$
separating $\Lambda_1$ from $\Lambda_2$ as that separation becomes large, tells us that distant sets of spins are conditionally
independent, if we condition on the spins in-between, when the model is in the finite-range class. To some extent, this justifies
the term finite-range, as it means that direct interactions between distant spins really are negligible in their effect;
indeed, the relative entropy that we bounded above for this case represents the distance from the state with those 
interactions dropped. 

In a general Gibbs state, the same would not be true if we did not condition on the spins
in the region separating $\Lambda_1$ from $\Lambda_2$, however, in a {\em pure} state it would hold without the conditioning,
because of the correlation decay property that characterizes pure states (see e.g.\  Georgii \cite{georgii_book}, Ch.\ 7). 
Hence in a Gibbs state
$\Gamma$, the mutual information $I({\cal S}_{\Lambda_1};{\cal S}_{\Lambda_2}|{\cal A})_\Gamma\to0$ in the limit, 
and then from a Markov chain argument like those above, applied to ${\cal A}\to{\cal S}_{\Lambda'^c}\to{\cal S}_{\Lambda_1}$
for $\Lambda_1\subseteq \Lambda'$ both finite, and $\Lambda'^c$ far distant from $\Lambda_1$, 
we find that 
\be
I({\cal S}_{\Lambda_1};{\cal S}_{\Lambda_2})_\Gamma\to I({\cal S}_{\Lambda_1};{\cal A})_\Gamma
\ee 
in the limit $\Lambda_2\to\Lambda'^c$ and $\Lambda'\to\infty$. That is, the mutual information in $\Gamma$ between 
${\cal S}_{\Lambda_1}$ and the spins ${\cal S}_{\Lambda_2}$ in a very large region $\Lambda_2$ very distant from $\Lambda_1$ 
can serve as a proxy for the complexity of $\Gamma$, and this extends immediately to $K_\Gamma$ and $K_\rho$. 

There are long-standing controversies surrounding the nature of the SG phase in classical SGs. It is our general
goal to shed light on these matters. To relate these to complexity, it will be useful first to review briefly some main results
of NS \cite{ns96b,ns_rev}. First, a metastate, which is a probability distribution on Gibbs states that contains information 
about behavior in finite size, can be either trivial, that is it consists of a $\delta$-function on a single Gibbs state, or nontrivial, 
meaning it is spread (or ``dispersed'') over more than one Gibbs state. Next, we again discuss the form of the pure state 
decomposition of a Gibbs state drawn from a given metastate. If the Hamiltonian has a global symmetry, say under $O(m)$, 
as it can in the $m$-vector models (including the Ising $m=1$ case), for example, then first we note that the metastate 
constructions preserve the symmetry, and a Gibbs state drawn from the metastate will possess the full symmetry of the Hamiltonian. 

If spontaneous symmetry breaking (SSB) occurs in a low temperature phase, then the Gibbs state will have a decomposition into 
pure states, at least some of which will not be invariant under the symmetry, and those will map to other pure states under the 
action of the symmetry, resulting in sets of pure states that each form an orbit under the symmetry action. (An orbit is defined by 
the property that the group acts transitively on it, that is any point can be mapped to any other point by a symmetry. If there is 
no global symmetry, a symmetry orbit is of course a single point.) The invariance of the Gibbs 
state is then preserved because its decomposition is uniform on each of its symmetry orbits of pure states. Note that for the Ising case, 
a nontrivial orbit has exactly two pure states in it, while for $SO(m)$ symmetry, $m>1$, a nontrivial orbit must be a continuum
[and in a SG will be a copy of $SO(m)$]. Thus the Gibbs state must have a decomposition into more than one pure state if 
SSB occurs. But this consequence of SSB is not so interesting for our purposes. Hence we will consider a Gibbs state to be trivial 
if it consists of a single orbit under the symmetry action, that is it decomposes into either a single invariant pure state, or into a 
single orbit consisting of more than one pure state; otherwise it is nontrivial. 

Then as both the metastate, and a typical Gibbs state drawn from it, can be either trivial or nontrivial, there are in principle 
four possible combinations of cases (in this discussion we assume that all the Gibbs states have the same character in this sense, 
as seems plausible). 
A further distinction that arises in NS's work is that the cardinality of the set of Gibbs states that may be obtained as a sample 
drawn from the metastate, and of the set of symmetry orbits of pure states that occur in the pure state decomposition of a Gibbs state, 
not only could be either one or greater than one, but also could be either countable (i.e.\ either finite or countably infinite), or uncountable. 
(Again, we assume that the answer to the last question for the pure state decomposition does not depend on the Gibbs state.) 
NS \cite{ns07} proved a theorem that states that, when the Gibbs states are nontrivial and their decomposition is finite or countable, 
the metastate must be supported on an uncountable number of Gibbs states, and in particular must be nontrivial. NS later
showed \cite{ns09} that in fact, almost surely, a Gibbs state drawn from the metastate is either trivial or consists of infinitely many 
symmetry orbits, eliminating the finite case just mentioned (see also Ref.\ \cite{ad}). We should note that the results mentioned here
involve the use of translation-invariance, obtained by using periodic boundary conditions in finite size.

The complexities we have defined are useful as a way to further quantify the degree of dispersal of the metastate
and the number of pure states in the decomposition of a Gibbs state. The $\nu$-expectation of the average mutual
information between the spins in a window and the pure state in a Gibbs state, ${\bf E}I({\cal S}_{\Lambda_W};{\cal A}|{\cal G})$ 
(where ``average'' refers to the average over the Gibbs state), is the (expected) complexity ${\bf E}K_\Gamma(\Lambda_W)$
of a typical Gibbs state $\Gamma$, relativized to the window $\Lambda_W$; we will denote this more simply as 
${\bf E}K_\Gamma(W)$, and similarly for the others. Likewise, the $\nu$-expectation of the mutual information between the 
spins in a window and the Gibbs state, ${\bf E}I({\cal S}_{\Lambda_W};{\cal G})$, is the (expected) complexity 
${\bf E}K_\kappa(W)$ of the metastate $\kappa$, relativized to the window $\Lambda_W$. [We recall that the complexity of the MAS 
$\rho$ is the sum of these two complexities, ${\bf E}K_\rho(W)={\bf E}K_\Gamma(W)+{\bf E}K_\kappa(W)$. 
We will leave the $\nu$-expectation implicit in the remainder of this discussion.] We have seen in Sec.\ \ref{subsubmono} that 
each of the complexities increases monotonically with $W$, so we can consider how fast they grow as $W\to\infty$. Typical forms 
that we may expect are growth as a power law in $W$, as a logarithm (or more generally perhaps a power of a logarithm), or bounded 
and tending to a non-negative constant (possibly zero). As we have seen, the upper bounds generally take a power-law form, so they 
place upper bounds on the exponent of the power (in one dimension, the bound may instead be constant, ruling out any form 
of unbounded growth). A complexity that grows as a power of $W$ should correspond to uncountable cardinality, while logarithmic 
or similar growth may correspond to a countable infinity. Of course, finite cardinality implies bounded complexity, tending to a constant, 
but not conversely. In applying these remarks to the questions of pure-state decomposition, it would be necessary to remove the 
contribution to the complexity from SSB, which we have seen implies that the cardinality is that of the continuum in the case of 
breaking a continuous symmetry. We have not considered how to do that. [Such an infinity was discussed at $T=0$ in Sec.\ 
\ref{subsubinfinite}. At nonzero temperature, SSB would imply uncountable cardinality, and infinite complexity as $W\to\infty$, 
but for finite $W$ thermal fluctuations would render the complexity finite.]

In many systems, even some with disorder, one does not expect to find infinite complexity or an uncountable number of pure 
or Gibbs states in a physical (e.g.\ a metastate) construction, other than when there is breaking of continuous symmetry. Hence 
the main applications of these ideas and results may be in SG theory. We can illustrate these applications by using various scenarios 
or models of SGs with Ising spins. First, in the scaling-droplet (SD) theory \cite{bm1,macm,fh}, the metastate and the Gibbs state 
were assumed (implicitly or explicitly) to be trivial, and so the complexities are zero (after subtracting the $\ln 2$ due to SSB). 

Replica symmetry breaking (RSB), in its (now) standard interpretation and its 
presumed application to finite-range Gibbs state \cite{par79,par83,mpv_book}, involves nontrivial Gibbs states, and also a
nontrivial metastate \cite{ns96b,ns07,read14}. Although in RSB the number of pure states in a given Gibbs state is countably infinite, 
the complexity of the Gibbs states is finite (as shown explicitly in Ref.\ \cite{gm}), and so this is a case in which, as $W\to\infty$, 
${\bf E}K_\Gamma(W)$ tends to a positive finite limit, which is the expected entropy of $\{w_\alpha\}$ \cite{hr}. The complexity 
of the metastate ${\bf E}K_\kappa(W)$ should then grow as a power law. We note that the exponent for the growth of complexity 
of the MAS is the maximum of the exponents for the other two. In Ref.\ \cite{hr}, the complexity of the MAS was used to define an 
exponent $\zeta'$ by 
\be
{\bf E}K_\rho(W)\sim cW^{d-\zeta'}
\ee 
as $W\to\infty$, for some constant $c$. (The notation $\zeta'$ references another 
exponent $\zeta$ that was defined \cite{read14} using a correlation function in the MAS, and argued there to equal $\zeta'$. 
In RSB, $\zeta=4$ at zero magnetic field for $d>6$ \cite{read14}.) We have the bound $\zeta'\geq 1$ in short-range models, and 
$\zeta'\geq 2\sigma-1$ for the one-dimensional $p=2$ power-law model when $\sigma\leq 1$, and these are valid for $T>0$, 
and for $m$-vector as well as Ising models. (This bound was mentioned recently in Ref.\ \cite{jry}.) Hence in this case $d-\zeta'$ 
describes the growth of complexity of the metastate. 

Another scenario, introduced by NS \cite{ns96b,ns_rev} for spin-flip invariant Hamiltonians, is called chaotic pairs.
In this case the Gibbs states are trivial (a flip-related pair of pure states), while the metastate is nontrivial. The analogous case
without spin-flip symmetry has been called chaotic singles \cite{read14}. In any of these cases the metastate might not contain an
uncountable number of Gibbs states, or might even be finite. The random-field Ising ferromagnet \cite{aw} exhibits chaotic
singles behavior, apparently with just two pure states (``up'' and ``down''). A number of short-range SG models with spin-flip 
symmetry, which differ somewhat from those considered in this paper, were constructed by White and Fisher \cite{wf}, and 
appear to possess chaotic-pairs SG phases, with power-law growth of ${\bf E}K_\kappa(W)$.

The remaining member of the set of four classes of combinations of trivial and nontrivial would be a phase in which the metastate is 
trivial, and the (single) Gibbs state is nontrivial (with an uncountable pure state decomposition). A similar phase arose in 
infinite-range SG models, and was associated with a transition into a dynamically-frozen phase (breaking ergodicity) \cite{kirk}. 
The complexity ${\bf E}K_\Gamma(W)$ of the Gibbs state appeared to be extensive, and the metastate appears to be trivial. 
It was then argued \cite{ktw,bb} that in a short-range analog, those states are instead ``metastable'' states, not pure, and the 
actual state is a ``mosaic'' of regions of those states, resulting in a single pure Gibbs state in place of the dynamically-frozen phase 
(the state is thus not distinct from the high temperature phase). As stated previously \cite{hr}, it is not clear to us why such effects 
must lead to a single pure state, rather than to a distinct phase in which there is a single Gibbs state with subextensive 
${\bf E}K_\Gamma(W)$ with growth exponent $d-\zeta'$ and $1\leq\zeta'<d$; the latter phase would then be an example 
of the final class.

Clearly the use of quantitative complexities and their rates of growth sharpens the discussion of the universal properties 
of the phases beyond simply the trivial-nontrivial and countable-uncountable distinctions. At present, we know of no further rigorous 
results that would eliminate any of the possible behaviors discussed above. Future work might produce stronger upper bounds on 
complexity than those found here, which could eliminate some possibilities. Of course lower bounds would be of great interest also.


\section{Conclusion}

Let us begin by summarizing some highlights of this paper. For finite-range models (see Sec.\ \ref{sgmodels}), 
we defined three complexities 
(\ref{compgam}), (\ref{compkap}), and (\ref{comprho}), which are respectively the complexity of a (typical) Gibbs state, 
of a metastate, and of the MAS. The last of these is the sum of the first two by eq.\ (\ref{comprel}). Each of the complexities can
be relativized to a finite window $\Lambda_W$ of size $W$, giving finite quantities $K_\Gamma(W)$, $K_\kappa(W)$, and 
$K_\rho(W)$ that increase monotonically with $W$; these obey the same relation $K_\Gamma(W)+K_\kappa(W) = K_\rho(W)$. 
A Markov chain argument showed that each of the relativized complexities is bounded by the mutual information between 
the spins in $\Lambda_W$ and those in its complement $\Lambda_W^c$, inequality (\ref{Ilamblamc}). From this point of view, 
what is important is that each complexity involves mutual information with random variables effectively at spatial infinity, and 
the locality of the finite-range models [expressed as the DLR definition of a Gibbs state, see eq.\ (\ref{gibbsdef})] implies that the 
information must be transmitted through the spins in $\Lambda_W^c$; this gives a bound that does not depend on the nature 
of the variables at infinity. With this,  for $T>0$, we obtained bounds (\ref{boundfinal}) or more generally (\ref{boundfinal'}) on the 
expectation value of each complexity, which have been discussed further in Sections \ref{intro} and \ref{subsubdisc}.
In order to do this rigorously for the general finite-range models it was necessary to provide
proofs of the existence and nature of Gibbs states for the long-range cases, which was done in Appendix \ref{app:gibbs}. 
A crucial (and non-elementary) step there was the use of the joint lower semicontinuity of the relative entropy, (\ref{semicon}), 
to transport bounds from finite size to the infinite-size limit.  

A number of issues remain open for further study. The bounds on the expected complexity give us no sense of how large the statistical 
fluctuations of each complexity may be; as the complexities are not proportional to the volume of the window, the size of fluctuations 
is not obvious (though square root of the surface area is a possible bound in the short-range cases).  
Finally, the growth exponent for each complexity is presumably a universal property, independent of most details of the model, 
and independent of $T$ at least for $T>0$ and within a given phase. There are very few circumstances in which we can calculate 
explicitly the expected complexity, or its growth exponent, in a SG model; the exceptions are scenarios and certain special models 
[see Sec.\ \ref{subsubdisc}]. This question, which is a more quantitative form of the basic question of whether or not there are many 
pure states or Gibbs states, gets to the heart of the SG problem in finite-range models.

\acknowledgements

I am grateful to J. H\"oller, S. Jensen, and A.P. Young
for stimulating collaborations, to C.M. Newman, D.L. Stein, and V. Vu
for discussions, to A.C.D. van Enter for discussions and references, to A. Barron 
for information on the lower semicontinuity of relative entropy under weak$^*$ 
convergence, to J. Fr\"ohlich for a remark during a talk by the author,
and to J.L. van Hemmen for correspondence.
The author is grateful for the support of NSF grant no.\ DMR-1724923.

\begin{appendix}

\section{Gibbs states, metastates, and pure states for long-range models}
\label{app:gibbs}

Here we give, for the finite-range models, a somewhat detailed account of Gibbs [i.e.\ Dobrushin-Lanford-Ruelle (DLR)] states,
which will be defined more stringently here, and discuss metastates also. Some of the discussion will be more technical than that in
the main text. We discuss an extension of the DLR definition
of a Gibbs state for long-range mixed $p$-spin models, and show that, within a metastate construction, the states produced satisfy that 
definition. We focus on Ising spins, but for $m$-vector spins we make some parenthetical side remarks when the necessary 
changes are more than merely notation. Some of the constructions (in particular, part of the proof of Proposition 3) will be 
utilized in the main text also (without circularity). 

\subsection{Partial sums and infinite-size limits}

The definition of a Gibbs state was begun in Sec. \ref{subsubgibbs}. The discussion frequently involves partial sums like
\be
\lim_{|\Lambda|\to\infty}\sum_{X\subseteq \Lambda:\ldots}\cdots,
\ee
where $\cdots$ is a function of $X$ and there may be additional stipulations $\ldots$ on the $X$ included in the sum
(sometimes it appears as $X\subseteq\Lambda'$ instead of $\Lambda$). 
These will be appear throughout the discussion in this Appendix, and are defined as follows: we evaluate the sum 
for $\Lambda=\Lambda_n$ (or $\Lambda'=\Lambda_n$), where $(\Lambda_n)_n$ is a cofinal sequence of finite subsets of 
${\bf Z}^d$, that is for any $X$, there is an $n$ such that $X\subseteq \Lambda_n$, and $\Lambda_n
\subset \Lambda_{n+1}$ (strictly increasing); then the limit is just $n\to\infty$. (It may be unconventional to include
strictness of increasingness in the definition, but it will be convenient.) These conditions ensure that eventually all $X$ (or all $X$ 
satisfying additional conditions) are included in the sum. In the short-range case of the sum (\ref{energysum}), convergence 
is absolute, and so does not depend on the choice of the cofinal sequence $(\Lambda_n)_n$ for $\Lambda'$, which appears there. 
Indeed, the set $\cal X$ of finite subsets of ${\bf Z}^d$ is countable, so here we could use any enumeration of $\cal X$ to define
the sum, and by absolute convergence it always gives the same sum. In the sums in the following, in principle we must 
specify the cofinal sequence $(\Lambda_n)_n$ in the notation whenever some terms in the sum could be negative; 
if not, or when we have reason to believe that the choice of cofinal sequence makes no difference, we write simply 
$\Lambda$ or $\Lambda'\to\infty$ to show the use of an unspecified cofinal sequence.  [For $m$-vector or other models, 
there may be sums over interaction types $x$ as well as over $X$. When the sums over $x$ are infinite for some $X$, 
we would need a cofinal sequence of finite sets of $(X,x)$. Under the additional condition imposed in Sec.\ \ref{subsubgibbs}, 
this will not be necessary in practice.] When discussing the existence of the thermodynamic limit for thermodynamics, it is 
necessary to specify more precisely how a cofinal sequence $\Lambda_n\to{\bf Z}^d$; for those we will stipulate 
(without further comment) convergence ``in the sense of Fisher'' 
\cite{cg_book,vEvH_83}, which holds when $(\Lambda_n)_n$ is a cofinal sequence of hypercubes, for example.

\subsection{Gibbs states}

For the finite-range models in general, the limit of a sum of the form (\ref{energysum}), which we reproduce here and 
which is involved in the definition of a Gibbs state as in eq.\ (\ref{gibbsdef}), 
\be
\lim_{n\to\infty}\sum_{X\subseteq \Lambda_n:\Lambda \cap X\neq\emptyset}J_X s_X,
\label{energysumapp}
\ee
may not exist at all, and if it does, it could depend on the cofinal sequence $\Lambda_n$ that is used. Before 
addressing that issue, we make further progress
by noticing an equivalent form of the definition of a Gibbs state \cite{georgii_book,simon_book,bovier_book}, 
in which the normalizing factor in $p_{H'}$ in eq.\ (\ref{gibbsdef}) is removed. This is 
done simply by taking ratios for distinct $s|_\Lambda$, for given $s|_{\Lambda^c}$:
\be
p_{H'}(s'|_\Lambda)/p_{H'}(s|_\Lambda)=\exp (-(H'(s')-H'(s))/T);
\ee
where $s'|_{\Lambda^c}=s|_{\Lambda^c}$ and $H'=H'_\Lambda(s)$ was defined in eq.\ (\ref{H'def}). 
(This applies for $T\geq 0$; for $T=0$, it can be viewed as defined by 
the limit $T\to0^+$, and the exponential may have to be interpreted as $0$ or $\infty$, even in a finite-size system.) 
Here one sees that this ratio is independent of $\Lambda$, provided $\Lambda$ is finite and  contains $\{i:s_i'\neq s_i\}$
as a subset. Hence it is sufficient to consider $\Lambda$ as small as possible, that is, the case $\Lambda=\{i:s_i'\neq s_i\}$, in 
which $s_i$ is changed, $s_i\to-s_i=s_i'$, for {\em all} $i$ in $\Lambda$, and $s_i'=s_i$ for $i\in\Lambda^c$; we then have 
(formally, because we have not yet discussed convergence of the infinite sum)
\bea
H'(s')-H'(s) &=& H(s')-H(s)\\
&=&\sum_{X:X\cap \Lambda\neq\emptyset}(-J_Xs_X'+J_Xs_X)\\
&\equiv&2h_\Lambda(s)
\label{energysumapp2}
\eea
(of course, $s_X'=\prod_{i\in X}s_i'$), which defines the formal sum $h_\Lambda(s)$ up to the convergence issue.
[For $m$-vector spins, we must consider $s'_i$ differing from $s_i$ by an arbitrary rotation for each $i\in\Lambda$, 
not simply reversal, so $h_\Lambda(s)$ is also a function of those rotations; for later use, the rotations should be constants, 
independent of $s|_\Lambda$.] In this form, the DLR definition of a Gibbs state says that
\be
\frac{\Gamma(s'_{\Lambda}\mid s|_{\Lambda^c})}{\Gamma(s_{\Lambda}\mid s|_{\Lambda^c})}=\exp [-2h_\Lambda(s)/T]
\label{dlralt}
\ee
for $\Gamma$-almost every $s$ and for all $\Lambda$.
Given the ratios for all $\Lambda\subseteq \Lambda'$ ($\Lambda'$ finite), the general conditional probabilities 
$\Gamma(s_{\Lambda'}\mid s|_{\Lambda'^c})$ can be recovered (which involves finding the normalizing factor). 
This then gives the full alternate version of the preliminary definition of Gibbs states, if the values of $h_\Lambda(s)$ 
are known.

In this form, for $T>0$ it might be that, for some $s$, $h_\Lambda(s)$ should be viewed as taking one of 
the values $\pm\infty$, which would mean that one of the two conditional probabilities is zero for this $s$. (In fact, 
we will soon see that $-\infty$ cannot occur.) This means that, on a set of $s$ of nonzero $\Gamma$-probability, 
$s'$ cannot occur: the spins in $\Lambda$ cannot all be reversed, and at least one is ``locked'', or not reversible, when 
starting from $s$. In the short-range case, $h_\Lambda(s)$ 
converges absolutely for all $s$ (and so is finite, $|h_\Lambda(s)|<\infty$), with $\nu$-probability $1$; hence in 
this Appendix we mostly focus on the long-range case (most statements also apply to short-range, some with trivial proofs). 
Thus we arrive at the question for long-range models: do locked spins occur in the Gibbs states when $T>0$? Gandolfi, Newman, 
and Stein (GNS) \cite{gns} showed that in long-range models, there exist distributions for the spins (i.e.\ states) in which some spins 
are locked, however those constructions are, by their own account, unphysical as they involve $J$-dependent boundary conditions. 
They gave a modified definition of Gibbs states in which locked spins (in the above sense) do not occur. 
(A related construction of Gibbs states in 
Ref.\ \cite{zegarlinski_87} is more closely related to the specification and DLR equations in Sec.\ \ref{app:reform} below; the results 
show that, when $T$ is sufficiently high, the thermodynamic limit from finite size exists, and locked spins do not occur.) 
   
The idea contained in the definition of a Gibbs state that appears in the work of GNS is that spin configurations $s$ 
that produce difficulties with the sums $h_\Lambda(s)$ occur with $\Gamma$-probability zero (we will extend their 
definitions from $p=2$ to general mixed $p$-spin models). We will need a definition for the meaning of the infinite sum. 
GNS took it to be defined using partial sums along a cofinal sequence $(\Lambda_n)_n$ (fixed independently of $s$, 
$\Lambda$, and $J$) as follows: retaining the same notation as above, we define, for $\Lambda'$ another finite set,
\be
2h_\Lambda(s|_{\Lambda'}) =\sum_{X\subseteq\Lambda':X\cap \Lambda\neq\emptyset}J_X(-s_X'+s_X),
\ee
and then define $h_\Lambda$ by
\be
h_\Lambda(s) = \lim_{n\to\infty}h_\Lambda(s|_{\Lambda_n}) 
\ee
for finite sets $\Lambda$ and configurations $s$ if this limit exists as an extended real number, that is either real or $\pm\infty$.
(It is not obvious that this is the most appropriate procedure.) 
Thus, at this stage we could allow the possibility $h_\Lambda=\pm\infty$. 

We can also define a version of $h_\Lambda(s)$ when $T>0$ directly from eq.\ (\ref{dlralt}):
\be
2\widetilde{h}_\Lambda(s)\equiv-T\ln \frac{\Gamma(s'|_\Lambda\mid s|_{\Lambda^c})}{\Gamma(s|_\Lambda
\mid s|_{\Lambda^c})}
\ee
provided the right-hand side exists. (Alternatively, we can rewrite this as a formula for $\widetilde{h}_\Lambda(s)/T$, 
and then it is valid at $T=0$ also). In these terms, we would want to show (and eventually, we will show) both that $h_\Lambda(s)$ 
exists and that it equals $\widetilde{h}_\Lambda(s)$ for $\Gamma$-almost all $s$. Here, however, as the idea is to consider 
$\widetilde{h}_\Lambda(s)$ as a random variable that depends on $s$, and $s$ is drawn from some distribution
$\Gamma(s)$, $\widetilde{h}_\Lambda(s)=-\infty$ would imply that, conditionally on $s|_{\Lambda^c}$, the spins 
are locked in configuration $s'$, and $s$ is forbidden. But, just as for elementary probability theory of discrete variables, 
$p(a|c)$ does not vanish unless also $p(a,c)=0$, so in general the $\Gamma$-probability of the set of $s$ such that 
$\Gamma(s|_\Lambda\mid s|_{\Lambda^c})=0$ must be zero. Thus with $\Gamma$-probability $1$, $\widetilde{h}_\Lambda(s)$ 
cannot be $-\infty$, when it exists.
 
$h_\Lambda(s)$ [and $\widetilde{h}(s)$] is a generalization of a local effective magnetic field $h_i(s)$: by taking $\Lambda=\{i\}$, 
we define $h_i(s)$ by $h_{\{i\}}(s)=h_i(s)s_i$. For $p=2$ $p$-spin interaction models, $h_i$ was used by GNS, and in that case, 
for general $\Lambda$, $h_\Lambda(s)=\sum_{i\in\Lambda}h_i(s)s_i -2\sum_{i,j\in\Lambda}J_{ij}s_is_j$, where the last sum 
is finite. In this case, for fixed $s$, existence of $h_\Lambda(s)$ for all finite $\Lambda$ is equivalent to existence of $h_i(s)$ 
for all $i$. (Here we assumed that, when $h_\Lambda(s)$ exists, it cannot be $-\infty$.)

Now, using the GNS definition of $h_\Lambda(s)$ as the limit of partial sums, following GNS we define the 
{\em allowed configurations\/} as those $s$ for which  $h_\Lambda(s)$ exists, and is finite, for all finite $\Lambda$ 
(with the given $J$). (Configurations that are not allowed will be called non-allowed. Also, for $m$-vector models, we must 
say ``for all $s'|_\Lambda$'' as well as ``for all finite $\Lambda$''.)  Finally, again following GNS, we make 
\newline
{\bf Definition 1}: a Gibbs state at $T\geq 0$ is a probability distribution $\Gamma$ on $S$, for given $J$, with 
the following two properties: \newline
1) the distribution is supported on allowed configurations only, that is $\Gamma(\{\hbox{allowed $s$}\})=1$;
\newline
2) for all finite subsets $\Lambda$, the conditional probabilities are given by the Gibbs form
\be
\Gamma(s|_\Lambda\mid s|_{\Lambda^c})=p_{H'}(s|_\Lambda)
\ee
[or alternatively by the formula for the ratios, so that $h_\Lambda(s)=\widetilde{h}_\Lambda(s)$], for $\Gamma$-almost 
every $s$ as before. 
\newline
We comment that: a) the evaluation of the right-hand side in 2) is well-defined 
as a consequence of property 1); b) this characterization of a given $\Gamma$ as a Gibbs state obviously depends on the definition
of $h_\Lambda$, and might depend on the choice of the sequence $(\Lambda_n)_n$ when partial sums are used;  
c) for $T>0$, vanishing 
conditional probabilities (i.e.\ $h_\Lambda(s)=+\infty$) for $s'$ given $s|_{\Lambda^c}$ do not occur, so there are no 
locked spins. While this certainly holds in the short-range models of this paper and in other short-range models, the 
motivation in general for this part of the definition may not be obvious at this stage. We will see that it is satisfied 
in the finite-range models considered here; d) the definition makes sense at $T=0$ as well as $T>0$, by taking $T\to 0^+$ 
as above. In this case, 2) implies that conditionally on $s$, for any finite set of sites, reversing those spins does not 
decrease the energy: $h_\Lambda(s)\geq 0$ for all $\Lambda$, and 1) implies that this energy change is well defined and finite. 
We call an allowed configuration $s$ a {\em ground state} if $h_\Lambda(s)\geq 0$ for all $\Lambda$; then the definition 
implies that, for $T=0$, $\Gamma(\{\hbox{ground states}\})=1$ (i.e.\ all configurations that occur are ground states). 
This is clear physically.

We note that at $T=0$, for non-degenerate $J$ a Gibbs state might be supported on a unique ground state 
(which would be a pure state), or if spin-flip symmetry is present, on the flip-invariant mixture of a pair of flip-related 
ground states (each of which is pure). A Gibbs state could also be a more general mixture of such states, implying non-uniqueness 
of ground states (or of ground state pairs). For degenerate $J$s, as is well known (for example, for $p=2$ interactions only, 
with $J_{ij}=\pm 1$ for all $\{i,j\}\in{\cal E}$, the set of nearest neighbor pairs), there could be extensive entropy at $T=0$, and 
individual spin configurations drawn from such a Gibbs state would still be ground states, but pure states would not be
supported on a single allowed configuration, because a single configuration would not be a Gibbs state.
Thus it would be a mistake to think that the general such Gibbs state necessarily possesses an extensive complexity of 
pure states (cf.\ Ref.\ \cite{palmer}). 

If $h_\Lambda(s)$ is defined in the manner of GNS as the limit of partial sums, then the allowed configurations
are defined without reference to temperature. We note that whether or not a series (with given coefficients $J_X$)
converges is a {\it tail event} in $s$, that is, the convergence (or not) of $h_\Lambda(s)$ is unaffected if $s_i$ is changed 
on a finite set of $i$. (It is also a tail event in $J$, similarly.) For similar reasons, the allowed configurations form 
a dense set in $S$ in the product topology, and so do the non-allowed (in the long-range case), because a base of open sets for $S$ 
is given by the collection of sets that are a product of open sets for a finite collection of $i$, times a copy of 
$S_0$ for all other sites \cite{chung_book,breiman_book,billingsley_book2,royden_book}, and so any non-empty 
open set contains both allowed and non-allowed configurations (in the long-range case). 
Turning to measure-theoretic properties, a $T=\infty$ Gibbs state would be the uniform distribution on $s$. That suggests 
that under that distribution, almost every configuration is allowed: for a configuration drawn from the uniform 
distribution, the sum $h_\Lambda(s)$ would converge (in the sense of partial sums along a cofinal sequence) 
with probability one. This is reminiscent of the textbook ``random signs'' problem of a series with independent 
uniformly random $\pm1$ factors in each term \cite{chung_book,breiman_book}, and indeed, for $p=2$, $h_i(s)$ 
reduces to exactly that form. In that case the series converges almost surely if and only if $\sum_j J_{ij}^2$ 
converges, which holds $\nu$-almost surely when the convergence condition holds. We may expect similar 
results for the more general mixed $p$-spin case for the uniform distribution on $s$. 
For general $T<\infty$, a state will not be absolutely continuous with respect to the uniform $T=\infty$ state,
and so it is not ruled out that it is not a Gibbs state under the above definition. Nonetheless, one would expect
that cancellations between terms, due to spins far apart being almost independent, could make the series converge
for typical configurations drawn from the state, and that is the type of result at which we are aiming. (See also
Refs.\ \cite{fz1,zegarlinski_87} for results at high $T$.)

Of course, so far we have only made a definition; the question now is whether the physically-relevant states
of finite-range SGs are Gibbs states in the sense of Definition 1. To address this, we will examine the states
produced by the metastate construction. 

\subsection{Metastates}

For the metastate construction, we follow AW \cite{aw} and NS, Ref.\ \cite{ns97} or \cite{newman_book} 
(see also Ref.\ \cite{bovier_book}), and sketch the main steps. First, we consider finite-size models on finite 
$\Lambda_{n'}'$ for a cofinal 
sequence $\Lambda_{n'}'\subset {\bf Z}^d$ indexed by $n'$ [not necessarily the sequence $(\Lambda_n)_n$ that might be 
used in defining allowed configurations]. Given $J$, we construct a finite-size Gibbs state $\Gamma^{n'}=
p_{H_{\Lambda_{n'}'}}(s|_{\Lambda_{n'}'})$, where $H_{\Lambda}=-\sum_{X\subseteq \Lambda} J_Xs_X$ for any
$\Lambda\in{\cal X}$, and $T\geq0$. $\Gamma^{n'}$ is a probability distribution on $s|_{\Lambda_{n'}'}$, and depends 
only on $J$ in $J|_{{\cal X}(\Lambda_{n'}')}$; it can be viewed as a consistent system of marginal distributions 
$\Gamma^{n'} = (\Gamma^{n'}(s|_\Lambda))_{(\Lambda,s|_\Lambda)}$ for $s|_{\Lambda}$, for all finite subsets 
$\Lambda\subseteq\Lambda_{n'}'$. Then for the joint distribution of the bonds and the state [that is, 
for the distribution of the pair $(J,\Gamma)$], by tightness and sequential compactness arguments 
\cite{chung_book,breiman_book,billingsley_book2}, there is a subsequence $n'_k$ of $n'$ such that all the 
final-dimensional marginal distributions for 
$(J|_{{\cal X}(\Lambda_{n'_k}')}, \Gamma^{n'_k})$ have $k\to\infty$ limits that form a consistent family of marginals; these define 
a joint probability distribution $\kappa^\dagger$ on $(J,\Gamma)$, where $\Gamma(s)$ (or $\Gamma_J(s)$ to show 
its dependence on $J$) is a probability distribution for $s$ in infinite size \cite{billingsley_book2}. [That is, 
$\left(J|_{{\cal X}(\Lambda_{n'_k}')}, \Gamma^{n'_k}\right)\to(J,\Gamma)$ as $k\to\infty$ in the sense of convergence in distribution, 
also known as convergence in law, or as vague or (in functional analysis) weak$^*$ convergence of the joint probability 
distributions for these random variables. The definition of weak$^*$ convergence for probability distributions $p_n(y)$, $p(y)$ 
on a space $Y$ is that $p_n\to p$ in the weak$^*$ sense as $n\to\infty$ if for all bounded continuous functions $f(y)$, 
$\int_Y f p_n\to\int_Y fp$.  
In the present case there is no need for a proof of tightness, because the
space of probability distributions $\Gamma(s)$ on $S$ is compact, while the marginal $\nu(J)$ for the bonds
is certainly tight.] The marginal distribution for $J$ is the original $\nu(J)$, and the 
conditional distribution on $\Gamma$ given $J$ is the AW metastate $\kappa$ (or $\kappa_J$ to show its 
dependence on $J$), or $\kappa(\Gamma)$ if we (probably inaccurately) imagine that it can be 
described by a probability density on states $\Gamma$. In other words, 
we can draw from the metastate $\kappa$ a (random) distribution or state $\Gamma(s)$ for the spins $s$. 

In this step, 
for the case of the finite-range mixed $p$-spin models there is almost no change in the argument, compared with the references; 
we note that the set $\cal X$ of all finite subsets of ${\bf Z}^d$, which indexes $J$, is countable, so that the spaces involved 
remain separable as $n'\to\infty$, allowing the argument to go through. As mentioned already, $\kappa^\dagger$ and $\kappa$ 
may not be unique: in addition to the choice of the sequence $(\Lambda_{n'}')_{n'}$ of finite sizes, a choice of a subsequence 
may have been required in order to obtain a limit. For notation, we will write formally $\kappa^\dagger=\nu\kappa$, 
which when the distributions can be represented by densities can be interpreted literally as $\kappa^\dagger(J,\Gamma)=
\nu(J)\kappa_J(\Gamma_J)$, and similarly when we also use $\Gamma$ to take expectations over $s$ we write 
$\kappa^\dagger\Gamma$ which (for densities) means $\kappa^\dagger(J,\Gamma)\Gamma(s)=\nu(J)
\kappa_J(\Gamma_J)\Gamma_J(s)$. Conditioning that on $J$ gives $\kappa_J(\Gamma_J)\Gamma_J(s)$,
which summed over $\Gamma$ gives by definition $\rho_J(s)$, the metastate-average state (MAS). 
This completes the construction of a metastate.

\subsection{Proof that the states are Gibbs states}

Next, we want to argue that the states $\Gamma$ (and not only the finite-size $\Gamma^{n'}$) drawn from a metastate
are Gibbs states in the sense of Definition 1; this is where, in comparison with Refs.\ \cite{aw,ns97,newman_book}, 
additional work is required. We recall that a cofinal sequence $(\Lambda_n)_n$ will be used; its form, which will 
depend on the distribution $\nu(J)$ (and is arbitrary for the short-range case) but will be independent of 
$J$, $\Gamma$, and $s$, and of the construction of $\kappa^\dagger$ [i.e.\ it is not necessarily the same as 
$(\Lambda_{n'}')_{n'}$ used in the construction of $\kappa^\dagger$], will be determined in the course of the proof. 
\newline
{\bf Theorem 1}: In the finite-range mixed $p$-spin models with a translation-invariant distribution 
(as defined in Sections \ref{sgmodels} and \ref{subsubgibbs}), for $T>0$ a state $\Gamma$ drawn from a metastate 
$\kappa$ is a Gibbs state (as in Definition 1), $\kappa^\dagger$-almost surely.
\newline
{\bf Proof}: The proof of Theorem 1 will take the remainder of this subsection, and follows immediately from the forthcoming 
Propositions $1$, $2$, and $3$; in those propositions, the hypotheses of Theorem 1 remain in force. QED.
\newline
{\bf Proposition 1}: For each finite $\Lambda$, the limit $\lim_{\Lambda'\to\infty}\Gamma(s|_\Lambda\mid s|_{\Lambda'-\Lambda})=
\Gamma(s|_\Lambda\mid s|_{\Lambda^c})$ exists $\kappa^\dagger\Gamma$-almost surely.
\newline
{\bf Proof}: (The proof is essentially standard \cite{aw,ns97,newman_book}, but included for completeness.) 
First, for $T\geq 0$, we consider the conditional probabilities 
$\Gamma(s|_\Lambda\mid s|_{\Lambda'-\Lambda})$ where $\Lambda$, $\Lambda'$ are both finite, and 
$\Lambda\subseteq\Lambda'$. [Strictly, one should say the conditional probabilities $\Gamma(s|_\Lambda\mid 
{\cal F}_{\Lambda'-\Lambda})$, where ${\cal F}_{\Lambda'-\Lambda}$ is the $\sigma$-algebra generated by 
$s|_{\Lambda'-\Lambda}$.] We note that, like any conditional probabilities, these are viewed as random variables, 
due to their dependence on $s|_{\Lambda^c}$, with distribution induced from $\Gamma(s)$. We fix $\Lambda$ 
and let $\Lambda'$ increase along any cofinal sequence $(\Lambda_n)_n$; then define $P_n=(\Gamma(s|_\Lambda\mid 
s|_{\Lambda_n-\Lambda}))_{s|_\Lambda}$, which is a sequence of random vectors, where each vector has components 
indexed by $s|_\Lambda$. (For $m$-vector models, this is better viewed as a regular conditional probability distribution 
on $s|_\Lambda$ \cite{breiman_book}.) Then the standard properties of conditional expectation \cite{chung_book,breiman_book}
imply that we have
\be
{\rm E}\,\left(P_{n+1} \mid s|_{\Lambda_n-\Lambda}\right) =  P_n;
\ee
in other words, taking the $\Gamma$ average over the spins in $\Lambda_{n+1}-\Lambda_n$ just produces the conditional
probability conditioned on $s|_{\Lambda_n-\Lambda}$. This means the conditional probabilities $P_n$ form a
martingale. (More formally, letting ${\cal F}_n={\cal F}_{\Lambda_n-\Lambda}$, the $\sigma$-algebras form a filtration 
${\cal F}_n\subseteq{\cal F}_{n+1}$, used to define the martingale; as $n\to\infty$, ${\cal F}_n\to{\cal F}_\infty$, 
the $\sigma$-algebra generated by $s|_{\Lambda^c}$ \cite{chung_book,breiman_book}.) The martingale convergence theorem 
\cite{chung_book,breiman_book} then tells us that, as $n\to\infty$, the conditional probabilities $P_n$ tend 
$\Gamma$-almost surely to a limit $P=(\Gamma(s|_\Lambda\mid s|_{\Lambda^c}))_{s|_\Lambda}$. As the notation 
suggests, the limit is independent of the sequence $(\Lambda_n)_n$ (or filtration ${\cal F}_n$) that was used in this 
part of the construction. This concludes the proof of Proposition 1. QED.

By Proposition 1, the conditional probabilities have limits as $\Lambda'\to\infty$, and hence $\widetilde{h}_\Lambda(s)/T$ 
is well-defined as an extended real number, and in fact for $T>0$ $\widetilde{h}_\Lambda(s)>-\infty$ 
$\kappa^\dagger\Gamma$-almost surely, as explained already.

Next we will prove
\newline
{\bf Proposition 2}: For $T>0$ and with $\kappa^\dagger\Gamma$-probability $1$, $\widetilde{h}_\Lambda(s)<\infty$.
\newline
{\bf Proof}: First take the average 
\be
{\bf E}_{\kappa^\dagger}{\rm E}\, \ln \frac{\Gamma(s)}{\Gamma(s')} ={\bf E}_{\kappa^\dagger}D[\Gamma(s)||
\Gamma_{\theta_\Lambda}(s)]
\ee
of the logarithm of the ratio of probability distributions, or of corresponding conditional probabilities, which give the
same result when conditioned on $s|_{\Lambda^c}$. (Here and in similar 
expressions below we abuse notation slightly on the left-hand side; it is an average over, not a function of, $s$, $s'$, just 
as the right-hand side is.) We have identified this with a relative entropy, as follows. 
If we consider the transformation $s\to s'$, where $s_i$ and $s'_i$ differ only in the finite set $\Lambda$, and express it
as $s'=\theta_\Lambda s$ where $\theta_\Lambda$ stands for the operation of reversing all spins in $\Lambda$
(or more general rotations for $m$-vector models and so on), we can instead consider this as a change in the state 
from $\Gamma(s)$ to $\Gamma_{\theta_\Lambda}(s) = \Gamma(s')$. 
(If $\Gamma$ is a Gibbs state, then $\Gamma_{\theta_\Lambda}$ is a Gibbs state in which $H$ has been transformed 
to $H_{\theta_\Lambda}(s)=H(\theta_\Lambda s)$. This uses the symmetry property of the uniform reference measure
on $S_0$ at each site that was mentioned in Sec.\ \ref{sgmodels}; no symmetry of $H$ is assumed. The symmetry also implies that
the following marginals are equal, namely $\Gamma_{\theta_\Lambda}(s|_{\Lambda^c})=\Gamma(s|_{\Lambda^c})$,
and hence it makes no difference whether we use the relative entropy of the distributions or of their conditionals
on $s|_{\Lambda_c}$ or a subset thereof.) 
Then the above ${\rm E}$ expectation is indeed, as displayed, the relative entropy of $\Gamma$ relative to 
$\Gamma_{\theta_\Lambda}$, with expectation using $\kappa^\dagger$ over the states $\Gamma$ and bonds $J$. 
This shows at once that ${\rm E}\, \widetilde{h}(s)/T\geq0$, 
including for the case $T=0$. Further, it can be shown \cite{pinsker_book} that, 
in the thermal average ${\rm E}\, \ln [\Gamma(s)/\Gamma_{\theta_\Lambda}(s)]$, the contribution of the negative part 
of $\ln [\Gamma(s)/\Gamma_{\theta_\Lambda}(s)]$ is $>-\infty$, which in particular again implies that 
$\widetilde{h}_\Lambda(s)/T>-\infty$ $\kappa^\dagger\Gamma$-almost surely. 

To arrive at an upper bound, we first recall the results for relative entropy of the Soviet authors, who included Gelfand, 
Kolmogorov, Yaglom, Perez, Dobrushin, and Pinsker, as described in Pinsker's book \cite{pinsker_book}. The relative entropy 
can be defined in general by partitioning configuration space $S$ into regions, finding the probability assigned 
to each region by each of the two probability distributions, and calculating the relative entropy for a partition using 
the formulas for discrete probabilities; then the relative 
entropy is the supremum, over partitions, of the relative entropy for a partition. In the present 
case we can define partitions by assigning each $s$ to the corresponding $s|_{\Lambda'}$, which indexes the regions, 
giving the relative entropies ${\rm E}\, \ln [\Gamma(s|_{\Lambda'})/\Gamma_{\theta_\Lambda}(s|_{\Lambda'})]$. 
(For continuous spins, the latter relative entropies must themselves be defined as the supremum over partitions
of $S_{\Lambda'}$ into a discrete set.) As 
$\Lambda'\to\infty$, the relative entropy associated to such a partition tends to the one we need. In fact, for all 
$\Lambda'$, ${\rm E}\, \ln [\Gamma(s|_{\Lambda'}) /\Gamma_{\theta_\Lambda}(s|_{\Lambda'})]$ is non-negative and increases 
monotonically to the limit \cite{pinsker_book}. (The monotonicity for relative entropy follows from the chain rule, similarly to
the argument in Section \ref{subsubmono}.)
These facts allow the use of the monotone convergence theorem \cite{royden_book}, so the 
limit can instead be taken after the $\kappa^\dagger$ expectation.
Thus we have
\bea
{\bf E}_{\kappa^\dagger}{\rm E}\, \ln \frac{\Gamma(s)}{\Gamma_{\theta_\Lambda}(s)}
&=&{\bf E}_{\kappa^\dagger}\lim_{\Lambda'\to\infty}{\rm E}\, \ln \frac{\Gamma(s|_{\Lambda'})}
{\Gamma_{\theta_\Lambda}(s|_{\Lambda'})}\\
&=&{}\lim_{\Lambda'\to\infty}{\bf E}_{\kappa^\dagger}{\rm E}\, \ln \frac{\Gamma(s|_{\Lambda'})}
{\Gamma_{\theta_\Lambda}(s|_{\Lambda'})}
\eea 
Here the first equality 
is from Ref.\ \cite{pinsker_book}, and the second 
is from the monotone convergence theorem. 

Now, using the second line, the expectation of the relative entropy for finite $\Lambda'$ can be upper bounded, using 
finite-size systems. For fixed $\Lambda'$, the thermal $\rm E$ average is a continuous function of $\Gamma$, and 
the $\kappa^\dagger$ expectation of that function calculated in finite size (i.e.\ its disorder average) tends to the infinite size 
version ($\kappa^\dagger$ average) by definition of convergence in distribution. In a finite system on $\Lambda'_{n'}$, 
where $\Lambda\subseteq \Lambda'\subseteq\Lambda'_{n'}$, we have, using notation from Section \ref{subsubbounds} 
and this Appendix (here ${\rm E}^{n'}$ is thermal expectation in $\Gamma^{n'}$), 
\bea
\lefteqn{{\bf E}\,{\rm E}^{n'}\,\ln \frac{\Gamma^{n'}(s|_{\Lambda'})}{\Gamma^{n'}(s'|_{\Lambda'})}=}\quad&&\non\\
&=&\frac{1}{T}{\bf E}\,{\rm E}^{n'}\, \left[-\left(F_{H-H_{\Lambda'}}(s|_{\Lambda'}) - F_{H-H_{\Lambda'}}(s'|_{\Lambda'})
\right)\right.\non\\
&&\left.{} + 2h_\Lambda(s|_{\Lambda'})\right]
\eea
The free energy difference would be zero if all interactions that involve spins in both $\Lambda$ and $\Lambda'^c$ were 
set to zero. Then by methods similar to those for finite-size bounds in Sec.\ \ref{subsubbounds}, in particular the bound (\ref{bound2}), 
we can obtain an upper bound 
at given $n'$, which under the convergence condition (\ref{cvgce_cond}) is finite, including for $n'\to\infty$. 

Next we wish to 
pass to the $n'\to\infty$ limit, in which the marginal states $\Gamma^{n'}(s|_{\Lambda'})$, 
$\Gamma^{n'}(s'|_{\Lambda'})$ converge in distribution (weak$^*$ convergence of the distribution) to limits $\Gamma(s|_{\Lambda'})$, 
$\Gamma(s'|_{\Lambda'})$, and we would like the $n'\to\infty$ limit of the bound to still hold for the expected relative entropy of 
$\Gamma(s|_{\Lambda'})$, $\Gamma(s'|_{\Lambda'})$. 
Such a limit of relative entropy under weak$^*$ convergence may not even exist, even when the probabilities are discrete
as for the case of Ising spins. However, we can make progress by first using the following general result \cite{posner}: on a 
fixed complete separable metric space, the relative entropy of probability distributions is ``jointly lower semicontinuous'' under 
weak$^*$ convergence, that is, if as $n'\to\infty$, $P_{n'}\to P$, $Q_{n'}\to Q$ in the sense of weak$^*$ convergence, then 
the relative entropy obeys
\be
D[P||Q]\leq\liminf_{n'\to\infty} D[P_{n'}||Q_{n'}]
\label{semicon}
\ee
(see Ref.\ \cite{royden_book} for $\liminf$ and for semicontinuity). Thus the relative entropy of a pair of states, such as
$\Gamma^{n'}(s|_{\Lambda'})$, $\Gamma^{n'}(s'|_{\Lambda'})$, is a lower semicontinuous function on pairs of states
in the appropriate topology on the space of such pairs. Next, any lower semicontinuous function $f$ 
on a compact metric space can be obtained as the pointwise limit of a pointwise-increasing sequence of bounded continuous 
functions $f_n$, that is $f_{n+1}\geq f_n$ for all $n$. Suppose then that $(f_n)_n$ is such an increasing sequence of bounded continuous 
functions of a pair of marginal probability distributions on $S_{\Lambda'}$ ($f_n$ independent of $n'$ for all $n$) that tends to the relative 
entropy; as relative entropy is non-negative, we can also assume that $f_n\geq0$ for all $n$. First applying $f_n$ to
$\Gamma^{n'}(s|_{\Lambda'})$, $\Gamma^{n'}_{\theta_\Lambda}(s|_{\Lambda'})$, we take 
$\lim_{n'\to\infty} {\bf E} f_n={\bf E}_{\kappa^\dagger} f_n$ by weak$^*$ convergence (along the subsequence $n'_k$ as always), 
which obeys the same $n'\to\infty$ limit of the upper bound as the right-hand side of eq.\ (A16), and then the $n\to\infty$ 
limit exists by the monotone convergence theorem, and gives the desired upper bound on ${\bf E}_{\kappa^\dagger} 
D[\Gamma(s|_{\Lambda'})||\Gamma_{\theta_\Lambda}(s|_{\Lambda'})]$.
Then taking the $\Lambda'\to\infty$ limit,  the expression is bounded 
above by $C_2T^{-2}\sum_{X:X\cap\Lambda\neq\emptyset}{\rm Var}\,J_X$ for a constant $C_2>0$ (the terms in the bound 
involving sites in both $\Lambda$, $\Lambda'^c$ go to zero in this limit), which for $T>0$ is finite. We conclude that 
${\bf E}_{\kappa^\dagger}{\rm E} \,\widetilde{h}_\Lambda(s)<\infty$.

As ${\rm E}\, \widetilde{h}(s)\geq0$, the fact that ${\bf E}_{\kappa^\dagger}{\rm E}\,
\widetilde{h}_\Lambda(s)$ is finite implies that ${\rm E}\, \widetilde{h}(s)<\infty$ $\kappa^\dagger$-almost surely.
(This result can also be viewed as saying that $\Gamma$ is absolutely continuous with respect to $\Gamma_{\theta_\Lambda}$,
$\kappa^\dagger$-almost surely, and also holds {\it vice versa} similarly.) Then as the contribution of the negative part of 
$\ln [\Gamma(s)/\Gamma_{\theta_\Lambda}(s)]$ is $>-\infty$, the contribution of the positive part must be $<+\infty$, 
and so, for $T>0$, $\widetilde{h}_\Lambda(s)<\infty$, $\kappa^\dagger\Gamma$-almost surely.  
This concludes the proof of Proposition 2. QED.

In the proof of the following Proposition 3, we will use the notion of an extended metastate. The meaning of extending
a metastate is as follows. It is sometimes desirable to carry additional information, other than the $J$ and the state
$\Gamma^{n'}$, from finite size systems through to the limit of infinite size. From the probability distributions in finite size, 
this is carried out as convergence in distribution starting from finite-dimensional marginal distributions, exactly as for 
the basic construction of  $\kappa^\dagger$ (in either AW or NS versions). If some $\kappa^\dagger$ is already given, 
then a (sub-) sequence $(n_k')_k$ was used, and (assuming tightness holds for the distributions of the additional data) 
one can find a subsequence of $(n_k')_k$ along which the distribution on the higher-dimensional data, as well as 
on $J$, $\Gamma$, converges, to obtain an extended distribution $\kappa^\dagger_{\rm ext}$, which conditioned on 
$J$ gives an extended metastate $\kappa_{\rm ext}$. As the sequence of finite sizes used is a subsequence of that 
used to obtain $\kappa$, the marginal distribution of the extended metastate on only the states $\Gamma$ is the 
original metastate $\kappa$. It is in this sense that $\kappa_{\rm ext}$ is an extension of $\kappa$ (or similarly for 
$\kappa^\dagger_{\rm ext}$ and $\kappa^\dagger$). Our use of extended metastates is inspired by the excitation 
metastate used by NS, and also by the ``natural'' metastate used in Ref.\ \cite{read18} to handle problems
similar to the present ones. The form of the extended metastate that we use here will be explained during the proof, 
but we emphasize that the statement of the Proposition does not involve the extension.
\newline
{\bf Proposition 3}: There is a cofinal sequence $(\Lambda_n)_n$ such that, for $T>0$, for all finite $\Lambda$, and 
for $\kappa^\dagger\Gamma$-almost every $s$,
\be
\widetilde{h}_\Lambda(s)=\lim_{n\to\infty} h_\Lambda(s|_{\Lambda_n}).
\ee
\newline
{\bf Proof}: The basic idea behind the proof is to compare $\ln \Gamma(s|_\Lambda\mid 
s|_{\Lambda^c})$ with the same quantity evaluated for a corresponding state $\Gamma_{(\Lambda,\Lambda_n)}$ 
in a system in which the interactions that connect spins in $\Lambda$ with those in $\Lambda_n^c$ are set to zero. 
In the long-range cases, that is an infinite set of bonds, and in trying to bound 
the effect of that change by the methods of this paper, which involve expectation over those bonds, sooner or later one 
reaches an expectation of an infinite sum of terms corresponding to those bonds, and the use of bound (\ref{nonGibp})
would be the next step. However, in general it is not clear if the expectation can be taken term by term. In order to render
the series a finite sum, we can consider a finite size system (as was done in the proof of Proposition 2). Then we
can use the metastate construction to take the limit, so that $\Gamma(s)$ is drawn from the metastate. However, it is not
clear if the basic metastate construction implies that the state $\Gamma_{(\Lambda,\Lambda_n)}$ with truncated 
interactions also exists in the limit. To deal with that, we introduce the truncated-interaction metastate,
an extension of the metastate $\kappa$, as follows. 

In finite size $\Lambda'_{n'}$, we construct for given $J$ the state $\Gamma^{n'}$, and also the states 
$\Gamma_{(\Lambda,\Lambda')}^{n'}$ for pairs of finite sets $\Lambda$ and $\Lambda'$, $\Lambda\subseteq\Lambda'$, 
in which all terms in the Hamiltonian that connect $\Lambda$ and $\Lambda'^c$ are dropped. Clearly, it
is sufficient to do this for $\Lambda$ and $\Lambda'\subseteq \Lambda_{n'}'$. 
It would be sufficient to use for $\Lambda'$ only the members of the eventual cofinal sequence $\Lambda_n$
(which will be characterized later), but use of all $\Lambda'$ (eventually without restriction, except $\Lambda\subseteq\Lambda'$) 
would work just as well. We then take the limit in distribution as $n'\to\infty$ of the joint distribution of 
$J|_{{\cal X}(\Lambda_{n'}')}$, $\Gamma^{n'}$, and the array $(\Gamma^{n'}_{(\Lambda,\Lambda')})_{(\Lambda,\Lambda')}$ 
in the same way as was done to obtain $\kappa^\dagger$, by choosing a subsequence of the sequence $(n'_k)_k$ used there, 
and call the resulting distribution $\kappa^\dagger_{\rm ext}$. 
As the finite subsets of ${\bf Z}^d$ are countable, the space of all these variables is again separable,
allowing the procedure to work. We denote the resulting truncated-interaction metastate by $\kappa_{\rm ext}$.

Then in finite size, we define
\be
\Delta={\rm E}\,\ln\frac{\Gamma^{n'}(s)}{\Gamma^{n'}_{(\Lambda,\Lambda')}(s)}
\ee ($s=s|_{\Lambda'_{n'}}$) which, being a relative entropy, is non-negative.  In the same way as before, 
we can derive bounds
\be
{\bf E}\Delta \leq\frac{C_3}{T^2}\sum_{X\in{\cal X}:X\cap\Lambda\neq\emptyset,X\cap\Lambda'^c
\neq\emptyset}{\rm Var}\,J_X,
\ee
in which we took $n'\to\infty$ on the right-hand side, and $C_3>0$ is a constant. As these bounds 
are finite under the convergence condition (\ref{cvgce_cond}), and independent of $n'$, they also apply to the 
limit, and so hold for the $\kappa^\dagger_{\rm ext}$ expectation of the relative entropy of
the states in infinite size, which are denoted $\Gamma$, $\Gamma_{(\Lambda,\Lambda')}$.
To be more careful here, we can instead begin with the relative entropy of the two marginal distributions 
for $s|_{\Lambda''}$ for $\Lambda''\subseteq\Lambda'_{n'}$ in finite size, which is less than that above as discussed 
in the proof of Proposition 2, so obeys the same bound. Let $n'\to\infty$ with $\Lambda''$ fixed (the lower semicontinuity 
property must be used again here; see the proof of Proposition 2), and then $\Lambda''\to\infty$ 
which, again as discussed in the proof of Proposition 2, gives the ($\kappa^\dagger_{\rm ext}$ expectation of the) 
desired relative entropy of $\Gamma$, $\Gamma_{(\Lambda,\Lambda')}$, and the upper bound again 
still holds. 
A similar bound applies when the thermal average is taken in $\Gamma_{(\Lambda,\Lambda')}$,
and also for the relative entropy of $\Gamma_{\theta_\Lambda}$, $\Gamma_{(\Lambda,\Lambda')\theta_\Lambda}$.
The finite upper bound implies that, $\kappa^\dagger_{\rm ext}$-almost surely, the two states 
are mutually absolutely continuous for any pair $\Lambda\subseteq\Lambda'$. (We can do the same for the relative entropy
of $\Gamma_{(\Lambda,\Lambda')}$ relative to $\Gamma_{(\Lambda,\Lambda'')}$, so any such 
pair are also mutually absolutely continuous.) 

Call the right-hand side of the upper bound $V'$.
We notice that 
\be
V'\leq \sum_{i\in\Lambda} V'_i
\ee
where $V'_i$ is $V'$ for $\Lambda=\{i\}$.
Hence $V'\to0$ as $\Lambda'\to\infty$ (through some cofinal sequence) for all $\Lambda$ if and only 
if the convergence condition (\ref{cvgce_cond}) holds. 
As relative entropy is non-negative, this implies that the  $L^1(\kappa^\dagger_{\rm ext})$-norm
of the relative entropy tends to zero, and hence the latter also tends to zero in probability. 

Next we will show that there is a sequence of sets $\Lambda'$ along which the relative entropy tends to zero 
$\kappa^\dagger_{\rm ext}$-almost surely. This subsequence is what we will take for $\Lambda_n$. 
We will describe such a sequence in detail to show that its properties are independent of $\Lambda$.
We use hypercubes $\Lambda_L$ of side $L$, $L$ an odd integer, centered at the same point in ${\bf Z}^d$ 
for all $L$; this sequence is clearly cofinal. Here we will take the center of all the cubes to be $i=0$, 
the origin (${\bf x}_0={\bf 0}$). Consider the sequences of $V'_i=V'_i(L)$ for each $i$, where 
$\Lambda'=\Lambda_L$, indexed by $L$ increasing through the odd integers; each sequence $(V'_i(L))_{L}$ decreases to zero. 
We intend to use $V'_0$ to define the sequence $(\Lambda_n)_n$. We need to show that $[V'_i(L)-V'_0(L)]/V'_0\to0$ as $L\to\infty$. 
This is straightforward, using the invariance of ${\rm Var}\,J_X$ under translations of $X$. We can show that
$|V'_i(L)-V'_0(L)|\leq V'_0(L-|{\bf x}_i|)-V'_0(L+|{\bf x}_i|)$, and as $\Lambda$ is fixed, $|{\bf x}_i|$ 
is fixed and bounded for all $i\in\Lambda$. From basic calculus the last difference is eventually much smaller than 
$V'_0(L)$. Then $V'$ is less than of order $|\Lambda|V'_0$, or we can say that for any $\varepsilon>0$, $V'(L)\leq (1+\varepsilon)
|\Lambda|V'_0(L)$ for sufficiently large $L$ (which depends on $\Lambda$ and $\varepsilon$). 

Now we use (the simplest form of) Chebyshev's inequality, that is, if $X$ is a non-negative random 
variable and $t>0$, then ${\bf P}[X\geq t]\leq{\bf E}[X]/t$; we apply this to the probability ${\bf P}_{\kappa^\dagger_{\rm ext}}$
in the truncated-interaction metastate. From this we obtain that, for any $\varepsilon\geq0$, for sufficiently large $L$ 
\be
{\bf P}_{\kappa^\dagger_{\rm ext}}[\Delta(L)\geq t]\leq \frac{(1+\varepsilon)|\Lambda|V'_0}{t}.
\ee
Now, for each $n$, define $L_n$ to be the smallest $L$ such that $V'_0(L)\leq 1/n^2$ 
for all $L\geq L_n$, and $\Lambda_n=\Lambda_{L_n}$. Then 
\be
\sum_n {\bf P}_{\kappa^\dagger_{\rm ext}}[\Delta(L_n)\geq t] \leq \frac{(1+\varepsilon)|\Lambda|\sum_n \frac{1}{n^2}}{t}
\ee
(at least for the tail of the sum at large $n$ on both sides), and the sum converges, so
\be
\sum_n {\bf P}_{\kappa^\dagger_{\rm ext}}[\Delta(L_n)\geq t]<\infty.
\ee
Hence for any $\Lambda$, by the Borel-Cantelli lemma  (see Ref.\ \cite{chung_book}, p.\ 77), the 
$\kappa^\dagger_{\rm ext}$-probability that $\Delta(L_n)\geq t$ for infinitely many $n>0$ is zero, and as $t>0$ 
is arbitrary it follows that $\Delta(L_n)\to0$ as $n\to\infty$, $\kappa^\dagger_{\rm ext}$-almost surely.

Relative entropy that tends to zero implies that $\Gamma_{(\Lambda,\Lambda_n)}\to\Gamma$ as $n\to\infty$ in a sense 
we now describe. Relative entropy obeys the Csisz\'ar-Kullback-Kemperman  inequality for two probability distributions 
$P_1$, $P_2$ (defined on the same $\sigma$-algebra on the same space $S$),
\be
d_{\rm TV}(P_1,P_2)^2\leq 2 D[P_1||P_2],
\ee
where the total variation (or $L^1$-) distance (i.e.\ metric) between any two signed measures $P_1$, $P_2$ 
is defined as
\be
d_{\rm TV}(P_1,P_2) = \int_S |dP_1-dP_2|
\ee
(see e.g.\ Ref.\ \cite{ct_book} for a simple proof;
some authors define the total variation distance to be half of ours). The total variation $d_{\rm TV}(P,0)$ is the norm 
on the space of finite signed measures $P$ that arises when that space is viewed as the dual of the space $C(S)$ of bounded continuous 
functions on the compact configuration space $S$. It defines a topology on the space of probability measures on $S$ that is 
(much) stronger than the weak$^*$ topology. Convergence in $d_{\rm TV}$ implies that the probability distributions become 
equal in the limit. Thus we have $\Gamma_{(\Lambda,\Lambda_n)}\to\Gamma$ in total variation distance (and hence also in 
the weak$^*$ topology, so that correlation functions also converge), $\kappa^\dagger_{\rm ext}$-almost surely. 

Then, from Proposition 2 and its analog for $\Gamma_{(\Lambda,\Lambda')}$, we have mutual absolute continuity 
of all the probability distributions $\Gamma$, $\Gamma_{(\Lambda,\Lambda_n)}$, $\Gamma_{\theta_\Lambda}$, 
$\Gamma_{(\Lambda,\Lambda_n)\theta_\Lambda}$ for all $n$, $\kappa^\dagger_{\rm ext}$-almost surely. We note that, 
while absolute continuity is not a symmetric relation, mutual absolute continuity is an equivalence relation, and distributions that 
are mutually absolutely continuous have exactly the same null sets; therefore, we can here say simply $\kappa^\dagger_{\rm ext}
\Gamma$-almost surely, regardless of which of these distributions we are using. Then we can take the Radon-Nikodym (RN) 
derivative \cite{royden_book}, for example $d\Gamma/d \Gamma_{\theta_\Lambda}$ of $\Gamma$ with respect to 
$\Gamma_{\theta_\Lambda}$, which is a function of $s$ that, because of absolute continuity, exists 
$\kappa^\dagger_{\rm ext}\Gamma$-almost surely; this is the true meaning of the ratio $\Gamma(s)/\Gamma_{\theta_\Lambda}(s)$ 
in the definition of the relative entropy, when the two distributions are mutually absolutely continuous \cite{pinsker_book}. 
We do the same with $\Gamma_{(\Lambda,\Lambda_n)}$ with respect to
$\Gamma_{(\Lambda,\Lambda_n)\theta_\Lambda}$. For a collection of mutually absolutely continuous measures,
the RN derivatives of one with respect to another enjoy properties similar to those 
of derivatives of a collection of functions with respect to one another, in particular the chain rule \cite{royden_book}. As 
$\Gamma_{(\Lambda,\Lambda_n)}\to\Gamma$ and $\Gamma_{(\Lambda,\Lambda_n)\theta_\Lambda}
\to\Gamma_{\theta_\Lambda}$ as probability measures, the RN derivatives tend to the $n\to\infty$ limit 
$\kappa^\dagger_{\rm ext}\Gamma$-almost surely, for example $d\Gamma_{(\Lambda,\Lambda_n)}/d\Gamma\to 1$. 
Taking logarithms, we have 
\be
\ln \frac{d\Gamma_{(\Lambda,\Lambda_n)}}
{d\Gamma_{(\Lambda,\Lambda_n)\theta_\Lambda}}=2h_\Lambda(s|_{\Lambda_n})/T.
\ee
and from the definition above,
\be
\ln \frac{d\Gamma}
{d\Gamma_{\theta_\Lambda}}=2\widetilde{h}_\Lambda(s)/T.
\ee
It follows that $h_\Lambda(s|_{\Lambda_n})\to\widetilde{h}_\Lambda(s)$ as $n\to\infty$, 
$\kappa^\dagger_{\rm ext}\Gamma$-almost surely and for all $\Lambda$. As the final statement 
does not refer to $\Gamma_{(\Lambda,\Lambda_n)}$, it in fact holds $\kappa^\dagger\Gamma$-almost surely; 
the proof is complete. QED.

We note that, once a metastate $\kappa$ has been obtained, a $\Gamma(s)$ drawn from it 
is well defined as a (random) state (a distribution on $S$). 
Only the characterization of the allowed configurations, and of $\Gamma$ as a {\em Gibbs} 
state, involved the sequence $(\Lambda_n)_n$. The proof of Proposition 3 shows that there is considerable leeway 
in the choice of the sequence, 
for example the choice of the common center for the hypercubes is arbitrary, or the hypercubes could be replaced by other compact 
shapes, and so on. A configuration $s$ that is allowed under the definition using one sequence will, with $\Gamma$-probability 
one, also be allowed under the definition using {\em any} other sequence such that the steps in the proof can be carried through.

\subsection{Reformulation and pure state decomposition}
\label{app:reform}

The GNS definition of a Gibbs state seems acceptable, but if we also wish to use some notion of pure 
Gibbs states, which preferably should possess the same properties as in the short-range cases,
then some reformulation is required. The usual description of a Gibbs state 
\cite{georgii_book,simon_book,bovier_book} begins from the notion of a family of so-called specifications, 
which normally correspond to the same family of conditional probabilities
$\Gamma(s|_\Lambda\mid s|_{\Lambda^c})$ (for $\Lambda$ finite) with which we began here. The specifications
$\gamma=(\gamma_\Lambda(A\mid s|_{\Lambda^c}))_{\Lambda}$ should for each $\Lambda$ be defined for 
all values of $s|_{\Lambda^c}$; here in most general form, these are probability kernels, so they are both 
a probability distribution on sets $A$ of $s$, and a measurable function of $s|_{\Lambda^c}$, and
are assumed to be proper (see Ref.\  \cite{georgii_book}, Chapter 1). They are defined {\it a priori}, 
meaning without specifying an unconditional $\Gamma$ first; they are independent of such $\Gamma$. They can be 
defined abstractly as possessing the properties of conditional probabilities, and to be Gibbsian they should be related 
to the Hamiltonian $H'_\Lambda(s)$. (Somewhat similar probability kernels, but conditioned on $s$, also 
arise as the transition probabilities for the dynamics of our systems \cite{liggett_book}.) Then the conventional definition is
\newline
{\bf Definition 2}: A Gibbs state admitted by a specification $\gamma$ is a probability distribution on $S$ whose 
conditional probabilities for each finite $\Lambda$ are: 
\be
\Gamma(A\mid s|_{\Lambda^c}) =\gamma_\Lambda(A\mid s|_{\Lambda^c})
\ee
$\Gamma$-almost surely.
\newline
By properties of conditional probabilities, this also implies that a Gibbs state is mapped to itself by each of a family (indexed 
by $\Lambda$) of linear maps of measures defined by the specifications (the DLR equations) 
\cite{georgii_book,simon_book,bovier_book}:
\be
\Gamma(A)=\int \gamma_\Lambda(A\mid s|_{\Lambda^c})\Gamma(s)ds
\label{dlreqs}
\ee
for all $A$ and $\Lambda$.
(In dynamics, corresponding conditions involving the transition probabilities say that $\Gamma$ is a 
stationary state \cite{liggett_book}.) Now to handle the long-range case, we only have to define the specifications 
to be used. Our proposal is to use the (generalized) specification, which in spirit is Gibbsian, defined for all finite $\Lambda$ 
and all $s$ by
\be
\gamma_\Lambda(s|_\Lambda \mid s|_{\Lambda^c})=\left\{\begin{array}{ll}p_{H'}(s|_\Lambda)&\hbox{if $s$ is allowed,}\\
                                                                                              0&\hbox{otherwise.}\end{array}\right.
\ee
We use the term ``generalized'' here (but will drop it hereafter) because if $s$ is not allowed, $\gamma_\Lambda$ 
is not in fact a probability; it gives zero
for any set of $s|_\Lambda$! However, the set on which it is zero is a tail event. Thus, rather than modifying the definition
of a Gibbs state, we have extended the definition of what can serve as a specification.
We remark that a) once again, the definition makes sense because of the definition of allowed $s$; b) this definition 
is consistent with the conditional probabilities of the Gibbs states under part 2) of Definition 1, because for any set of 
configurations that has zero probability, the conditional can be defined arbitrarily; c) conversely, by eqs.\ (\ref{dlreqs}), 
Gibbs states under the current definition assign probability $1$ to the set of allowed $s$, as required by part 1) of Definition 1.
Hence, for these specifications, Definition 2 is equivalent to Definition 1. 

A conventional route to proving the existence of a decomposition of any Gibbs state into a mixture of
pure (i.e.\ extremal) states involves showing that the Gibbs states admitted by the specification $\gamma$ form a 
set ${\cal G}(\gamma)$, which as a subset of the space of probability measures on $S$ (with, say, the weak$^*$ topology) 
is clearly convex, and also closed. Closure is guaranteed if the maps of measures defined by the specification 
are continuous. That then gives a compact convex set of probability measures, and the Choquet theory of such 
sets \cite{phelps_book} leads to the desired results. In the present case, for the long-range models, 
continuity of the above $\gamma$ is not obvious. [Nor is it ``quasilocal''; see remark (2.22) in Ref.\ \cite{georgii_book}.]
However, Georgii's book \cite{georgii_book}, in particular sections 7.1 and 7.3, leading up to Theorem 7.26, 
characterizes the pure states and gives results of Dynkin and F\"ollmer that establish that any Gibbs state has a unique 
decomposition into pure states, without the use of any such topological properties of $\gamma$. That decomposition is 
then the starting point for the analysis of the complexity of Gibbs states discussed in this paper. 
In the present case, there are some $s$ for which $\gamma$ fails to be a probability distribution, but one can verify that 
the proofs of the main results from Chapter 7 of Georgii \cite{georgii_book} still go through  in this case 
with only minor modifications (in particular, formulas involving $\gamma$ that hold $\Gamma$-almost surely 
for Gibbs states $\Gamma$ are unaffected).

Finally, we should note that Ref.\ \cite{zegarlinski_87} uses a specification defined on the set of the allowed
configurations $s$ only, and shows for any $T>0$ that any state obtained as a limit from finite size (with boundary conditions) must 
satisfy the DLR equations (\ref{dlreqs}) (however, a proof there that the state puts measure $1$
on the set of allowed $s$ holds only at sufficiently large $T$). This does not seem to be sufficient for our purposes.

\section{Uniqueness of Gibbs states at $T>0$ in short-range case in one dimension}
\label{app:trivi}

In this Appendix we provide a short and fairly simple proof that at $T>0$ there is a unique Gibbs state
(a pure state) in any short-range mixed $p$-spin SG model in dimension $d=1$ that satisfies 
one simple condition. (Under the same conditions, this rules out a non-trivial metastate, and also 
rules out any phase transition that would imply a change in the number of pure states.) For the $p=2$ 
power-law model of Ising spins, arguments for similar results were given in earlier work 
\cite{khanin,kas,vEvH_85,cove} (see also Ref.\ \cite{vEf} for a similar result in the case of 
short-range $m$-vector models with $O(m)$ symmetry and $d\leq 2$). Ref.\ \cite{khanin}
gives a complete proof for $\sigma>3/2$. Following the proposal of Kotliar {\it et al.\/} that there would be no 
transition at $T>0$ for $\sigma>1$ \cite{kas}, van Enter and van Hemmen \cite{vEvH_85} employed 
a simple approach based on relative entropy to show the absence of spontaneous breaking of spin-flip symmetry
in that region, but their paper and Ref.\ \cite{vEf} were criticized for some technical issues in Ref.\ \cite{cove}. 
The latter \cite{cove} employs a very different approach and arrives at a full proof for $\sigma>1$
in a set-up using fixed-spin boundary conditions, but the proof is rather long and some may find it difficult. 
We note that similar results for short-range non-disordered spin systems, such as Ising ferromagnets, 
are well-known folklore (and for strictly short-range cases can be proved easily using a transfer matrix), 
and were proved rigorously in Ref.\ \cite{ruelle}; see also Ref.\ \cite{simon_book} (p.\ 303) and references therein for 
a simpler proof. The strategy of our proof is to show that
the relative entropy of two distinct pure states is bounded, which gives a contradiction; this is similar to that of Ref.\ 
\cite{vEvH_85}, but we implement it in a form that avoids some technical questions, using an upper bound 
exactly like those elsewhere in this paper. The statement in the Theorem is more general than in Refs.\ 
\cite{khanin,vEvH_85,cove}; in particular, other than existence of the first two
absolute moments, we do not use a condition on the tail of the distribution of $J_X$. 

The statement is
\newline
{\bf Theorem 2}: Consider a short-range SG model (as defined in Sections \ref{sgmodels} and \ref{subsubgibbs}) 
in $d=1$, with sites ${\bf x}_i=i\in{\bf Z}$, and $T>0$. If the bonds satisfy
\be
\sum_{X\in{\cal X}:X\cap{\bf Z}_-\neq\emptyset,X\cap{\bf Z}_+\neq\emptyset}{\rm Var}\,J_X < \infty
\label{domwall}
\ee
(where ${\bf Z}_-=\{i\leq0\}$ and ${\bf Z}_+=\{i>0\}$), then there is at most one pure Gibbs state, $\nu$-almost surely.
(For $m$-vector models with $m>1$, or other non-Ising cases, the sum ranges over $x$ as well as $X$, with no additional
conditions on $x$.)

Before starting the proof, we discuss some general points. First, as we consider only short-range cases, most of the 
technicalities of Appendix \ref{app:gibbs} will not be needed here; in particular, as we begin simply from pure states,
the metastate construction is not required, though it is useful in that it establishes that some Gibbs states actually exist.

Second, the fact that the Gibbs states considered in the proof are pure is not used until the end of the proof. We will need the
fact that distinct pure states are supported on disjoint sets of spin configurations \cite{georgii_book}, and so are mutually
singular. (This may be physically obvious if the supports of the distinct pure states are viewed as sets of configurations 
that are mutually inaccessible in dynamics, i.e.\ as ergodic components.) This behavior is the extreme opposite
of absolute continuity, and implies that the relative entropy of either with respect to the other must be infinite. 
If we examine the relative entropy of one with respect to the other for the marginal distributions in a finite region
(window), then (for Gibbs states) the result will be finite, but it will increase monotonically (see proof of Proposition 2
in Appendix \ref{app:gibbs}) to infinity as the size of the region expands until it eventually includes all the sites (i.e.\ 
along a cofinal sequence). That is, for any choice of a bound $M$, the relative entropy must eventually be greater than $M$ 
for all sufficiently large window sizes. Of course, in our case, the relative entropy will be a random variable, and the 
statements must be made probabilistically. We will 
also use in the next paragraph the fact that, in a pure state, connected thermal correlations of functions 
of spins in well-separated regions tend to zero as the separation goes to infinity (see Refs.\ 
\cite{georgii_book,simon_book,bovier_book} for the precise statement).

Third (some readers may prefer to skip this on a first reading), we will be interested in an expectation 
of a relative entropy, which is a thermal average, here in a pure state. The expectation involves two pure states 
as (some, at least, of) the $J$s are varied, and the question may arise whether we can do this here: can we be sure 
that we obtain the ``same'' two pure states as some of the $J_X$ are changed? (This is related to the concerns 
in Ref.\ \cite{cove}, that construction of pure states might require $J$-dependent boundary conditions, 
preventing naive manipulations of an expectation over all $J$.) In the present case, we will be interested
in the expectation ${\bf E}'$ over only the bonds $J_X$ such that $X$ has non-empty intersection with a finite interval, 
say 
\be
\Lambda_W=\{-(W-1)/2,-(W-1)/2+1,\ldots,(W-1)/2\}
\ee
($W>0$ odd); call that expectation ${\bf E}'_W$. Here, we discuss this question in general, for any 
dimension $d$. If we begin with a given Gibbs state, say $\Gamma$, for some given $J$, we can 
actually construct a corresponding Gibbs state $\Gamma'$ with other values of some of the $J_X$s, as follows. 
For a change in Hamiltonian $\Delta H$, the Gibbsian formulation suggests a definition of a perturbed state, 
such that expectation of any function of $s$ in $\Gamma'$ is related to that in $\Gamma$ by 
\be
\langle\cdots\rangle_{\Gamma'}=\frac{\langle \cdots e^{-\Delta H/T}\rangle_\Gamma}{\langle 
e^{-\Delta H/T}\rangle_\Gamma}
\ee
where $\langle\cdots \rangle_{\Gamma'}$ ( $\langle\cdots \rangle_\Gamma$) denotes expectation 
in $\Gamma'$ (respectively, $\Gamma$). If $\Delta H=-\sum_X\Delta J_X s_X$ has the form of a general 
mixed $p$-spin Hamiltonian, but includes nonzero terms for only a {\em finite} set of $X$, then 
the perturbed expectation can be expanded out in terms of averages in $\Gamma$ 
(with coefficients based on those in $\Delta H$) and the change in its value is finite. This fact extends to the 
case of $\Delta H$ containing an infinite number of terms, if $\Delta H$ also satisfies the absolute 
convergence condition for a short-range Hamiltonian like that which precedes eq.\ (\ref{sh_cvgce_cond}).
Namely, if we define 
\be
|| J ||_1\equiv \sum_{X}|J_X|
\ee 
[the $l^1$ norm on $J$; in models other than Ising, the sum must range over the pairs $(X,x)$ as before],
then using $|\Delta H|\leq||\Delta J||_1$ a sufficient condition is that $||\Delta J||_1$ be finite.
In that case, the infinite sum converges because the terms in its tail decrease sufficiently rapidly, and their effect on 
thermal averages converges also, by use of an easily-proved inequality such as
\be
\left|\frac{\langle f(s) e^{-\Delta H/T}\rangle_\Gamma}{\langle 
e^{-\Delta H/T}\rangle_\Gamma}- \langle f(s) \rangle_\Gamma\right| \leq (e^{2||\Delta J||_1/T}-1)\sup_s |f(s)|
\ee
where $f$ is any function of $s$ [cf.\ eq.\ (A.1.6) of AW \cite{aw}; this topic forms part of their discussion
of properties of metastates]. 
This means that (in models of the general form defined in Sec.\ \ref{sgmodels}), Gibbs states (in the weak$^*$ topology) 
are continuous functions of $J$ (in the topology determined by the norm $||\cdots||_1$) \cite{aw}. Further, if 
$\Gamma$ is a pure state, the asymptotic behavior 
of thermal averages is unaffected by such a change in the Hamiltonian, so $\Gamma'$ is also pure. This is 
not difficult to show if $\Delta H$ contains only a finite number of terms, by using the asymptotic decay 
of connected correlations in $\Gamma$ \cite{georgii_book}, and it extends 
to all cases in which $||\Delta J||_1<\infty$ by a simple approximation 
argument. In effect, the change in the Hamiltonian is only a local one, without detrimental long-range effects.  
In the case of interest, $\Delta H \propto H'_{\Lambda_W}(s)-H_{\Lambda_W}(s)$, and $||\Delta J||_1$ is finite $\nu$-almost surely 
for a short-range model. We will now adopt the corresponding perturbations of the 
original two pure states with which we began, and so view them as functions of the bonds in question.

{\bf Proof} of Theorem 2: Suppose that there are two distinct pure Gibbs states $\Gamma_\alpha$ and $\Gamma_\beta$. 
We will bound their expected relative entropy. Consider their marginal distributions on, without loss of generality, 
$\Lambda_W$. Form the relative entropy 
\be
D_{\alpha\beta,W}={\rm E}_\alpha\ln \frac{\Gamma_
\alpha(s|_{\Lambda_W})}{\Gamma_\beta(s|_{\Lambda_W})}
\ee
of $\Gamma_\alpha(s|_{\Lambda_W})$ with respect to 
$\Gamma_\beta(s|_{\Lambda_W})$, and take the ${\bf E}'_W$ expectation (for which, see the discussion before 
this proof). It can be bounded by the method that by now should be familiar, using the (formal) interpolating Hamiltonian
$H +(\lambda-1)(H'_{\Lambda_W}-H_{\Lambda_W})$ where $\lambda$ runs from zero to one, and the perturbation of
the states proportional to $\lambda-1$ can be handled as explained before this proof; we call the resulting pure states 
$\Gamma_\alpha^{(\lambda)}$, $\Gamma_\beta^{(\lambda)}$. We note that, at $\lambda=0$, 
$\ln \Gamma^{(0)}_\alpha(s|_{\Lambda_W})/\Gamma_\beta^{(0)}(s|_{\Lambda_W})=0$, as $\Lambda_W$ is 
decoupled from the rest of the system. Hence $\ln \Gamma_\alpha(s|_{\Lambda_W})/\Gamma_\beta(s|_{\Lambda_W})$ 
is equal to the integral $\int_0^1d\lambda$ of
\bea
\frac{d}{d\lambda}\ln  \Gamma^{(\lambda)}_\alpha(s|_{\Lambda_W})&=&-\frac{1}{T}\left(\left\langle 
H'_{\Lambda_W}-H_{\Lambda_W}
\right\rangle_{\alpha,s|_{\Lambda_W},\lambda}\right.\non\\
&&{}\left.-\left\langle H'_{\Lambda_W}-H_{\Lambda_W}\right\rangle_{\alpha,\lambda}\right),
\eea
minus the similar expression with $\beta$ in place of $\alpha$. [The notation here is similar to that in Sec.\ \ref{subsubbounds}, 
though the (conditional) thermal expectations are taken using $\Gamma_\alpha^{(\lambda)}$, as indicated.] We have 
to be careful about taking the expectation (doing integrals) term by term on the sum, because the method 
of returning to a finite-size system is not available here. There are in fact up to three integrals 
or expectations (namely those implied by ${\bf E}'_W$, ${\rm E}_\alpha$, and $\int_0^1 d\lambda$), 
and also the sum. If we replace each term $J_X s_X$ (or in some places a conditional ${\rm E}_\alpha$ or ${\rm E}_\beta$ 
average of such a term) by its absolute value, then the short-range condition (\ref{sh_cvgce_cond}) implies 
that the sum converges, and so by part of the Fubini-Tonelli Theorem \cite{royden_book} it is legitimate to carry out 
the integrals and sum in any order, and the result is finite. That proves that the integrand-summand is integrable, and 
so the other part of Fubini-Tonelli tells us that the integrals and sum of the original series can be carried out in any order 
also. Then we obtain:
\be
{\bf E}'_WD_{\alpha\beta,W}\leq C\frac{1}{T^2}\sum_{X:X\cap\Lambda_W\neq\emptyset,X\cap\Lambda_W^c\neq\emptyset}
{\rm Var}\,J_X
\ee
where $C>0$ is a constant, and of course also the same with $\alpha$ and $\beta$ interchanged. (The $m$-vector models
involve summation over $x$ as well as $X$.) The upper bound 
increases with $W$. As $W\to\infty$, it is finite if and only if the hypothesis (\ref{domwall}), 
which arises from the contribution of each of the two ends of the interval, holds. 

Going back to $W$ finite, the same bound 
applies for each $W$ if we take the conditional expectation over a larger set of bonds than those involved in ${\bf E}'_W$, say those 
for $X$ intersecting some finite $\Lambda_n$, $\Lambda_W\subseteq\Lambda_n$. As we take such a set larger and larger (i.e.\ 
$\Lambda_n\to\infty$ along a cofinal sequence), effectively removing the conditioning on more and more bonds, because 
of the upper bound (and positivity of the relative entropy) the backward martingale convergence theorem 
\cite{chung_book,breiman_book} tells us that, for each $W$, the limit of the conditional expectation exists as a random variable 
that is measurable with respect to the tail $\sigma$-algebra of $J$, and that it obeys the same bound above. We now work in this  
limit, where expectations are the full $\bf E$, and similarly for the distribution $\nu$ (strictly speaking, they are still 
conditioned on the tail $\sigma$-algebra of $J$).

If we consider the infinite sequence of relative entropies $D_{\alpha\beta,W}$ for all finite odd $W>0$,
and if the hypothesis (\ref{domwall}) holds, then using Chebyshev's inequality the family, indexed by $W$, 
of distributions (induced from $\nu$) of $D_{\alpha\beta,W}$ for finite $W$ is tight: no weight goes off to infinity 
as $W\to\infty$. Because $D_{\alpha\beta,W}$ is an increasing function of $W$, it tends either to a finite limit or to infinity 
as $W\to\infty$, so the preceding result shows that, $\nu$-almost surely, it tends to a finite limit:
the probability that the two states are mutually singular is zero. 
As the Gibbs states were assumed to be pure and distinct, this contradiction shows that any two pure states are in fact identical, so 
there is a unique pure state, and a unique Gibbs state. That concludes the proof. QED. 

We comment that in the $p=2$ $d=1$ power-law models both the short-range condition 
and condition (\ref{domwall}) imply that $\sigma>1$. In general, a comparison of the two conditions involves
the dependence of ${\rm Var}\,J_X$ on both $|X|=p$ and ${\rm diam}\, X>0$ (in one dimension, we can define the diameter
of $X$ as ${\rm diam}\,X=\max\{i:i\in X\}-\min\{i:i\in X\}$). 

An information-theoretic interpretation of the proof of Theorem 2 is that the short-range interactions in $d=1$
are not strong enough to convey to a finite window $\Lambda_W$ sufficient information concerning in which pure state 
the system is supposed to be; for pure states, that information (the relative entropy) would have to 
tend to infinity with $W$. 

An alternative, slightly weaker, statement of the result is that, for any two pure states in the decompositions
of two respective MASs, they must be equal, implying the triviality and uniqueness of the metastate and of the 
Gibbs states drawn from it. In this case the proof can use the expectation ${\bf E}_{\nu(\kappa_1w_1\times\kappa_2w_2)}$ 
over pairs of pure states in the decompositions of respective Gibbs states drawn from respective metastates $\kappa_1$ 
and $\kappa_2$ (where for $\tau=1$, $2$, $w_\tau=w_\alpha(\Gamma_\tau)$ denotes the decomposition of a Gibbs state 
$\Gamma_\tau$, say, drawn from $\kappa_\tau$, into pure states $\Gamma_\alpha$ for the given $J$), and over bonds. 

We emphasize that the proof holds for pure states in any $m$-vector model, not only those produced by applying 
$O(m)$ symmetry with an $O(m)$-invariant Hamiltonian, but also those produced by non-$O(m)$-invariant Hamiltonians. 
If we wish only to show the nonexistence of $T>0$ pure states that spontaneously break the global spin-flip symmetry 
[$O(1)={\bf Z}_2$] in the Ising case, for a flip-invariant Hamiltonian that contains terms with even $p$ only, 
then we can consider a pure state $\Gamma_\alpha$ and the state obtained by flipping all spins in $\Lambda_W$, 
as in Proposition 2 above, and closer to Ref.\ \cite{vEvH_85}. The argument is similar to before, and involves marginalizing 
to a region $\Lambda'$ containing $\Lambda_W$, and taking $\Lambda'\to\infty$, so essentially we bound 
${\bf E}{\rm E}_\alpha h_{\Lambda_W}(s)$ (see Appendix \ref{app:gibbs}), then finally take $W\to\infty$. 
In the resulting bound, the domain wall sum in condition (\ref{domwall}) is modified to include only terms 
indexed by $X$ such $|X\cap{\bf Z}_-|$ and $|X\cap {\bf Z}_+|$ are both odd, and there is no symmetry 
breaking if that sum is finite (thus when terms with $p>2$ are present, this condition is slightly less restrictive than 
the direct application of that found above to the present models). The same sum is also found if we extend to 
mixed even-$p$-spin models the proof for $d=1$ of an upper bound \cite{read18} that leads to a bound on 
the exponent $\theta$ in the scaling-droplet theory \cite{fh}; when the sum is finite, the scaling-droplet arguments 
predict that there is no transition that spontaneously breaks spin-flip symmetry at $T>0$.

Very similar proofs as for Theorem 2 and for the absence of symmetry breaking at $T>0$ hold for the case of non-random 
$p$-spin interactions \cite{ruelle} (without the need for any disorder average, of course), and reproduce the 
well-known classic result for only $p=2$ interactions, which has $|J_X|$ in place of ${\rm Var}\,J_X$ in the 
domain-wall sum, and which is then indeed a bound on the energy of a domain wall in the ground state. 
For the case of proving the absence of spontaneous breaking of continuous symmetry in short-range SGs 
in $d\leq 2$, an upper bound on the relative entropy stronger than one like that used above is required; see 
Ref.\ \cite{vEf}. For that case, there are also other arguments \cite{sy} in the style of the Bogoliubov 
inequality approach, and we will not pursue it here.

\section{Thermodynamic limit, convergence conditions, and infinite-range models}
\label{app:energy}

In this Appendix we briefly discuss some questions of the existence of the thermodynamic limit for thermodynamic properties, 
such as the free (and ground state) energy density, and some other
convergence conditions related to these; we also add some comments about infinite-range models. 
In this Appendix, we do not assume all of the conditions of Sections \ref{sgmodels} and \ref{subsubgibbs} on the distributions until 
later; instead, we generally assume only that the bonds $J_X$ are independent and (for simplicity) centered for {\em all} $X$.

The proof of the existence for $\nu$-almost every $J$  of the thermodynamic limit of the free energy per spin at $T>0$ 
in a finite-range model is in principle analogous to that of the strong law of large numbers \cite{chung_book,breiman_book},
if the free energy is roughly the sum of almost independent free energies of distinct regions of the system. A proof, 
following Ref.\ \cite{vEvH_83,cg_book}, is based on two ingredients, from which the result follows by a subadditive ergodic theorem. 
The ingredients are the subadditivity of the free energy when two or more disjoint finite parts are coupled together, and a lower bound 
on the expected free energy per spin that is uniform in the system size. For the case of independent, centered $J_X$s, subadditivity is 
straightforward to show for the {\em expectation of} the free energy \cite{cg_book}, while the proof of the lower bound \cite{vEvH_83} 
has been extended to this case in Ref.\ \cite{cg_book} (and references therein). We will show briefly how the latter proof can be 
improved and simplified using methods from the body of this paper. We note that Ref.\ \cite{zegarlinski_91} directly proves almost-sure 
existence of the limit under the same conditions that we discuss below.

First consider a finite system of sites in a set $\Lambda$. Using the Gibbs distribution, we consider $\ln \sum_s e^{-H(s)/T}=-F/T$,
introduce a factor $\lambda$ by replacing $H=H_\Lambda$ by $\lambda H$, and then consider the integral from 
$0$ to $1$ of the derivative with respect to $\lambda$. This gives (for Ising spins) the identity
\be
-\frac{1}{T}F=|\Lambda|\ln 2 +\frac{1}{T}\int_0^1d\lambda\,\sum_{X:X\subseteq\Lambda} J_X\langle s_X\rangle_\lambda,
\label{freeenid}
\ee
and then clearly this is bounded above by
\be
\leq|\Lambda|\ln 2 +\frac{1}{T}\sum_{X:X\subseteq\Lambda} |J_X|
\label{freeenabsbo}
\ee
for any $J=(J_X)_X$.
This upper bound on (minus) the free energy $F$ (divided by $T$) can be obtained directly: the first 
term is the maximum entropy, and the second is (minus) a lower bound on the internal energy $-\sum_X J_X\langle s_X\rangle$ 
(divided by $T$), which would be attained if every bond were satisfied (usually that is not possible).
Then for independent, centered bonds, a sufficient condition for the existence of the thermodynamic limit for the expected 
free energy per spin for $T>0$ is
\be
\lim_{\Lambda\to{\bf Z}^d}\frac{1}{|\Lambda|}\sum_{X:X\subseteq\Lambda}{\bf E} |J_X|<\infty.
\ee
For homogeneous distributions, we can also express this as
\be
\lim_{\Lambda\to{\bf Z}^d}\sum_{p\geq1}\,\frac{1}{p}\sum_{X\subseteq\Lambda:i\in X,|X|=p}{\bf E}|J_X|<\infty
\ee
for any $i$. For the models we defined in Sec.\ \ref{sgmodels}, if the set of $p$ that contribute to the sum is finite, 
this is equivalent to the condition (\ref{sh_cvgce_cond}) for the model to be short range, but not when the set of $p$ is infinite;
in that case the present condition is weaker. 

If we return to the identity (\ref{freeenid}) and take its ${\bf E}$ expectation, then for independent,
centered $J_X$, we can use inequality (\ref{nonGibp}), applied to each ${\bf E}\,J_X\langle s_X\rangle_\lambda$, to obtain
\be
-\frac{1}{T}{\bf E}F\leq |\Lambda|\ln 2 +\frac{1}{2T^2}\sum_{X:X\subseteq\Lambda} {\rm Var}\,J_X.
\label{freeenvarbo}
\ee
[As in Sec.\ \ref{subsubbounds}, for Gaussian bonds the same bound can be obtained by integration by parts, or otherwise 
\cite{cg_book}; for distributions that are not necessarily Gaussian, these bounds (\ref{freeenabsbo}) and (\ref{freeenvarbo}) 
are stronger than the corresponding ones in Sec.\ 3.4 of Ref.\ \cite{cg_book}. We also note that the variance of the Hamiltonian 
is given by the same sum,
\be
{\bf E} H_\Lambda^2 =\sum_{X\subseteq \Lambda} {\rm Var}\,J_X.
\ee
This is for the Ising case; for $m$-vector models, the equality should be replaced by $\leq$, and as usual the sum 
should range over $x$ as well as $X$.] 

If we now divide the bound (\ref{freeenvarbo}) by $|\Lambda|$, then the condition that the right-hand side be finite 
as $|\Lambda|\to\infty$ 
is another sufficient condition for the existence of the thermodynamic limit for the expected free energy density when $T>0$ 
\cite{zegarlinski_91,cg_book}. For homogeneous distributions, it reduces to
\be
\lim_{\Lambda\to{\bf Z}^d}\sum_{p\geq 1}\,\frac{1}{p}\sum_{X\subseteq \Lambda:i\in X,|X|=p} {\rm Var}\,J_X <\infty.
\ee
For the models we defined in Sec.\ \ref{sgmodels}, if the set of $p$ that contribute to the sum is finite, this is equivalent 
to the convergence condition (\ref{cvgce_cond}) for the model. If the set of $p$ is infinite, then the convergence condition 
implies this one, but not conversely. Hence we see that one can construct models on ${\bf Z}^d$ in which the free energy
density possesses a thermodynamic limit, but in which Gibbs states presumably do not exist, and there may be locked spins 
(cf.\ Appendix \ref{app:gibbs}). These models are not finite-range,
and neither are they infinite range in the sense defined in Sec.\ \ref{sgmodels}. They exist even when the number of terms 
(indexed by $X$) for which $i\in X$ and $|X|=p$ is {\em finite} for all $p$, simply because of a divergence of the sum over $p$ in 
(\ref{cvgce_cond}). 

The case of the infinite-range models runs parallel to the long-range models. Although they never have Gibbs states in 
a strict (DLR) sense, the same condition (\ref{cvgce_cond}), where now ${\rm Var}\,J_X$ for $p>1$ depends on $|\Lambda|$, 
implies that the molecular field on a given spin is finite 
including when $T\to0$, as one can see heuristically, for example by using a replica symmetric ansatz \cite{sk,gm}; a related 
function involving the overlaps arises in the Parisi formula for SK-type models (see e.g.\ Ref.\ \cite{pan_book}, where
again it is assumed that the sum in the convergence condition converges sufficiently rapidly). 
Hence there are also SK-type models that have a limit for the free energy density \cite{gt,cg_book}, but not for the 
molecular field, and which are thus more singular than what we called infinite-range models. An example of this phenomenon is
Derrida's random-energy model, when it is viewed as the $p\to\infty$ limit of the pure $p$-spin models \cite{derrida}. 
The scaling of ${\rm Var}\,J_X$ is such that the thermodynamic limit of the free energy per spin exists, which implies that the 
($p$ times larger) convergence condition sum tends to infinity. Hence it is not surprising that
in the thermodynamic limit of the random-energy or $p\to\infty$ $p$-spin model there is a low-temperature region 
in which the entropy per spin is zero \cite{derrida}, implying that the spins are (in effect) locked into a small number 
of configurations. We expect similar locking phenomena in other models that satisfy the weaker condition above 
but not the stronger condition (\ref{cvgce_cond}), including in the models 
on ${\bf Z}^d$ that we mentioned in the preceding paragraph.

Except in the case of short-range models, the bounds so far on the free energy per site are not effective at $T=0$.
However, the variance of the Hamiltonian per site enters a general lower bound on the expectation of the ground state 
energy $E_0(\Lambda)=\min_{s}H_\Lambda(s)$ in a finite system $\Lambda$ (and hence also of the internal energy 
at $T\geq 0$) for any SG of Ising spins in which the variance of $H_\Lambda$ is independent of $s=s|_\Lambda$, 
such as the mixed $p$-spin models in this paper; the bound does not seem well known in the physics literature. 
Here we assume the $J_X$s are independent, centered, and Gaussian. The bound is (see e.g.\ Ref.\ \cite{chatterjee_book})
\be
{\bf E}E_0(\Lambda)\geq -\sqrt{{\bf E}H_\Lambda^2}\sqrt{2|\Lambda|\ln 2},
\ee
so the expected ground state energy per spin has a finite limit if the variance per spin does. (If ${\bf E}H_\Lambda^2$ is not independent
of $s$, it can be replaced by its maximum to obtain the bound. Of course a similar bound applies to ${\bf E}\max_s 
H_\Lambda(s)$.) 
This bound can be used to simplify an argument in Ref.\ \cite{read18} (see Proposition 6 there) that locked spins do not occur 
in the $p=2$ model at $T=0$; that argument can be generalized to give the same statement for any finite-range {\em pure} $p$-spin 
model at $T=0$. 

The methods here and in Ref.\ \cite{cg_book} suffice to prove the existence of a limit for the expected free (or ground state) 
energy per spin under the conditions stated. A proof of almost sure convergence of the free (or ground state) energy per spin 
as in Refs.\ \cite{vEvH_83,cg_book} can be obtained if one can prove either the subadditivity of the free energy without taking 
the expectation \cite{vEvH_83}, or else that the free energy per spin concentrates at its expectation as the limit is taken 
\cite{cg_book}. For either of these, some additional conditions may be necessary, but that lies outside the scope of this paper. 
Ref.\ \cite{zegarlinski_91} proves almost-sure existence of the limit for $T>0$ without such additional conditions. 

\end{appendix}

\end{document}